\documentclass[useAMS,usenatbib]{mn2e}
\usepackage{graphicx}
\title[Contribution of a Disk Component to Single Peaked Broad Lines of  Active Galactic Nuclei]{Contribution of a Disk Component to Single Peaked Broad Lines of  Active Galactic Nuclei}

\author[E. Bon et al.]{E. Bon$^{1,2}$, L. \v C. Popovi\'c$^{1,2}$,
 N. Gavrilovi\'c$^{1,2,3}$, G. La Mura$^{4}$, E. 
Mediavilla$^{5}$\\
$^{1}${Astronomical Observatory,  Volgina 7, 11160 Belgrade 74, Serbia, email: ebon@aob.bg.ac.yu}\\
$^{2}${Isaac Newton Institute of Chile, Yugoslavia Branch}\\
$^{3}${Observatoire de Lyon, 9 Avenue Charles Andr\'e, Saint-Genis Laval Cedex, F-69561, France}\\
$^{4}${Department of Astronomy, University of Padova, Vicolo dell'Osservatorio, I-35122 Padova, Italy}\\
$^{5}${Instituto de Astrofisica de Canarias, Tenerife, Spain}\\}
\begin{document}

\date{Accepted 1988 December 15. Received 1988 December 14; in original form 1988 October 11}

\pagerange{\pageref{firstpage}--\pageref{lastpage}} \pubyear{2009}

\maketitle

\label{firstpage}

\begin{abstract}
We study the disk  emission component  hidden in the single-peaked Broad
Emission Lines (BELs) of Active Galactic Nuclei (AGN). We compare the observed broad lines from a
sample of 90 Seyfert 1 spectra taken from the Sloan Digital Sky Survey with simulated line profiles.
We consider a two-component Broad Line Region (BLR) model where an accretion disk and a surrounding non-disk region with
isotropic cloud velocities generate the simulated BEL profiles. The analysis is mainly based in
measurements of the  full widths (at 10\%, 20\% and 30\% of the maximum intensity) and of the
asymmetries of the  line profiles.  Comparing these parameters  for  the simulated and observed
H$\alpha$ broad lines, we { found} that the hidden disk emission { may } be present in BELs even
if the characteristic {of two peaked line profiles is}  absent. For the available sample of objects (Seyfert 1 galaxies with single-peaked BELs), our study indicates that, { in the case of the hidden disk emission in single peaked broad line profiles}, the disk inclination tends to be small (mostly $i<25^\circ$) and that the contribution of the  disk emission to the total flux should be smaller than the contribution of the surrounding region.

\end{abstract}

\begin{keywords}

galaxies: Seyfert, accretion, accretion disks, line: profiles.

\end{keywords}

\section{Introduction}

Modeling of the double-peaked Balmer lines has been used to study the emission gas kinematics of the
Broad Line  Region (BLR)  \citep[see e.g.][]{Perez88, Chen89a, Chen89b, EH94, EH03, Rod96,  Storchi97,
Liv97, Ho00, Shields00, Strateva03, Berg03, Stor03}. However, only a small fraction of Active Galactic
Nuclei --  AGN (3\%-5\%)  shows  clearly double peaked Broad Emission Lines (BELs) in their spectra
\citep{Strateva03}. 

According to the standard unification model \citep{urpad95} one can expect an accretion disk around a
supermassive black hole in all AGN. The majority of AGN with BELs, have only single peaked lines, but
this does not necessarily indicate that the contribution of the disk emission to the BELs profiles is
negligible. It is well known that  a face-on  disk also emits single peaked broad lines
\citep[see e.g.][]{Chen89b,Dum90,Koll02,Koll03}. Moreover, a Keplerian disk of arbitrary inclination
with presence of a disk wind can also produce single-peaked broad emission lines \citep{Mur95}.

In spite that most of the BELs are single peaked, there are other evidences like the detection of asymmetries and substructure (shoulders or bumps, for instance) in the line profiles that indicate the presence of a disk (or disk-like) emission 
\citep{Pop02,Koll03,Shap04}. { Also}, the study of the accretion rates in AGN  supports the
presence of a standard optically thick and geometrically thin disk \citep{Wang03}.  Moreover, the spectropolarimetric observations gave an evidence for the disk-like emission  \citep[rotational motion, see e.g.][]{Smith05}

To explain the complex morphology of the observed BELs shapes, different geometrical models have been
discussed \citep[see in more details][]{Sul00}. In some cases the BELs profiles can be explained only
if two or more kinematically different emission regions are considered   \citep[see e.g.][]{Rom96,
Pop01, Pop02, Pop03a, Pop04, Pop08, Pop09, Bon06, Bon08, Ilic06, Col06,Hu08}. In particular, the existence of a Very Broad Line
Region (VLBR) with random velocities at ~5000-6000 km/s within an Intermediate Line Region (ILR) has
also been considered to explain the observed BELs profiles \citep{CB96,Sul00,Hu08}.

In this paper we study the presence of the hidden disk emission in objects which show only one dominant
peak in their broad emission line profiles.  To do that we consider that the BLR has two kinematic
components; an accretion disk and a surrounding non-disk region.  With this model we compute emission
line profiles for different values of the model parameters. Then we compare the simulated profiles (specifically, their  widths and
asymmetries)  with the observational data. 

{ The aim of this paper is to discuss possibility that the disk geometry, at least partly, affects the complex BEL profiles, i.e. to try to constrain the BLR geometry that is  important for estimates of the AGN black hole masses and accretion rates  \citep[see e.g.][]{lam09}.}

The paper is organized as follow: in \S2 we describe the two component model of the BLR and perform
the numerical simulation. In \S3 we compare the simulations with available data. In \S4 we  discuss our results and in \S5 we outline our conclusions.

\section{Numerical simulations based on a two-component model}

\subsection{BLR geometry}

In the last  years, arguments supporting the presence of disk winds show ability to explain a number
of observed AGN phenomena such as the X-ray and UV absorption, line emission, reverberation results, 
some differences among Seyfert and other active objects  (like quasars or broad-line radio galaxies),
and the presence or absence of double-peaked emission-line profiles  \citep[see
e.g][]{Mur95,Mur98,Proga04}.   These results  support a model in that the BLR is { composed} from 
two kinematically distinct components, a disk and a
wind. Recently, \citet{Hu08} { confirmed} that the BLR is { probably} composed from two emission regions,  i.e.
a VBLR and a ILR component, { as it was earlier assumed in several papers \citep[see e.g.][etc.]{CB96,Sul00,Pop04}} .

Consequently, we assumed that the BELs can be kinematically divided  into  two components, one
 from  the VBLR (contributing to the wings) and  other from the ILR (contributing to the core). Of
course, { one can assume different geometries for both emitting regions} \cite[see e.g.][{ and also Appendix B in this paper}]{Pop04}, but here we will assume
that the VBLR is coming from an accretion disk, and the ILR from { an} additional region, which surrounds the
disk, and has an isotropically distributed  random velocity. Note here that the kinematics of a wind would imply radial velocities (logarithmic profile), more than isotropic (Gaussian profile), but we assume Gaussian profile as a first approximation. The scheme of the assumed model is presented in Fig. 1.

\begin{figure}\centering\includegraphics[width=8 cm]{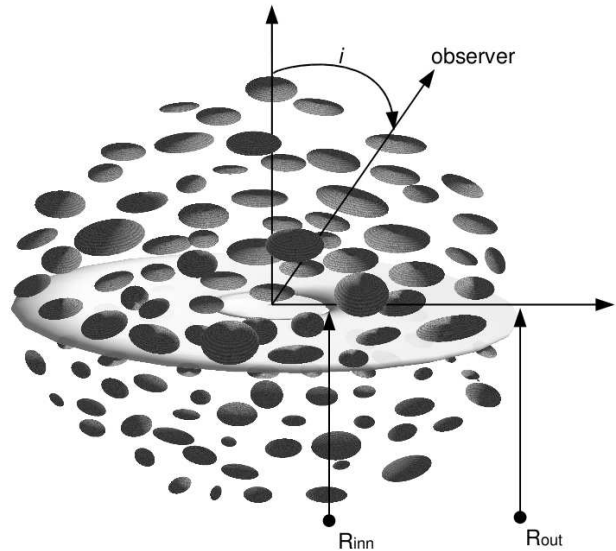} \caption{The scheme of the BLR composed from two geometrically different components: accretion disk and  clouds randomly distributed around the disk.} \end{figure}

The local broadening  ($\sigma$) and shift ($z_l$) of each disk element have been taken into
account as in \cite{Chen89b}, i.e. the $\delta$ function has been replaced by a Gaussian function: 

$$\delta\to\exp{{{(\lambda-\lambda_0-z_l)^2}\over{2\sigma^2}}},\eqno(1).$$

We express the disk dimension in gravitational radii ($R_g=GM/c^2$, $G$ being the gravitational
constant, $M$  the mass of the central black hole, and $c$  the velocity of light).

On the other hand, we assume that the additional emission region can be described by a surrounding
region with {an isotropic velocity distribution}, i.e. the emission line profile generated by this region can be described by a Gaussian function with broadening $w_s$ and shift $z_s$. Thus, the whole line profile can be described by the relation:

$$I_{}(\lambda)=I_{{d}}(\lambda)+I_s(\lambda),\eqno(2)$$
where $I_{{d}}(\lambda)$,  and $I_s(\lambda)$ are the emissions of the
relativistic accretion disk and   the non-disk  region,
respectively.

\subsection{Parameters for the disk and surrounding region}

As it was earlier noted in \citet{Pop04}, this two component model
can fit the line profiles of  BELs, but  is too open to constrain the physical parameters. First of all, the disk
model includes many parameters (the size of  the emitting region,
the emissivity and inclination of the disk, the velocity dispersion of the
emitters in the disk, etc.). 
 Therefore, in order
to do numerical tests, one  needs to introduce some constraints and
approximations. 

It seems that the {{parameter $\sigma$ of the}}  Doppler broadening of the non-disk region
and  { {parameter $\sigma$ corresponding to the broadening of the random motion in the disk model}} are connected \citep[see][]{Pop04,Bon06}. As a first
approximation we assume that the random velocities in the disk and in 
the non-disk region are the same. { {So}} here we consider {{a parameter $\sigma=1000$\ km/s for both, the Doppler broadening  of the non-disk regions as well as the $\sigma$ in the model of the disk profile}} \citep[also, see][]{EH03}).\footnote{ In Appendix A we give simulations  of different $\sigma$ for non-disk region}

On the other hand, we considered a wide
range of disk parameters but with several constraints:

i) The disk inclination affects the  emission obtained from
the disk. The observed flux from the disk
($F_{d}$) is proportional to the disk surface ($S_{d}$), as

$$F_{d}\sim S_{\rm eff}\sim S_{d}\cdot \cos(i),\eqno(3)$$
where $i$ is the inclination, and $S_{eff}$ is the effective disk emitting
surface, therefore, one cannot expect a high
contribution of the disk emission to the total line profile for a near edge-on projected  disk.

ii)  As far as the plasma inside 100 gravitational radii from the central black hole is very hot, one
cannot expect emission of the low ionized lines in this part of the disk. Then we limit the inner radius
to  $R_{inn}>100$ R$_g$. Consequently, the model given by  \citet{Chen89b} can  be
properly used, i.e. it is not necessary to include a full relativistic calculation
\citep[as e. g. in][]{jov08}.

iii) The emissivity of the disk as a function of radius, $r$, is given by $\epsilon=\epsilon_0r^{-p}.$
Since the illumination is due to a point source  radiating isotropically, located at the center of the
disk, the flux in the outer disk at different radii should vary as $r^{-3}$ \citep{EH94}. { We note
here that  this is indeed the way how the incident  flux varies, but not necessarily the way in which
lines respond to it  \citep[see e.g.][]{Dum90,RB99,EH03}. However, the power index $p\approx 3$ can be
adopted  as a reasonable approximation at least for H$\alpha$ \citep{EH03}. Also, 
\citet{caowang05} indicate that $p$ is in the range from 2 to 3. Moreover we simulated the influence of the emissivity ($p$)
to the disk line profile (Fig. 2) and found that it only slightly affected the normalized disk line profiles. Therefore the
assumption of $p=3$ can be accepted for purposes of this work.

\begin{figure}
\centering
{\includegraphics [height=6cm,width=8cm] {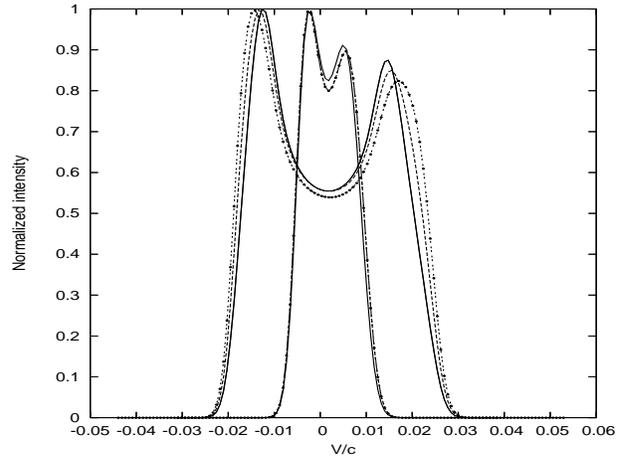}}
 \caption{Simulated disk profiles for different values of the 
emissivity: p=2 (solid line) p=3 (dashed line) and p=4 (doted-dashed line), 
for two inclinations $i=$10 (narrower lines) and 30 degrees (broader line profiles). The inner and outer radius are 
taken as:  $R_{inn}=$500 Rg and   $R_{out}$=1500 Rg. The maximum intensity is scaled to one.} \label{fig:Simulation+}
\end{figure}

iv) Previous estimations  of  the double-peaked AGN emission lines
\citep[see e.g.][]{Perez88, Chen89a, Chen89b, EH94, Rod96, Storchi97,
Ho00, Shields00, Strateva03, Berg03, Stor03, EH03} show that the typical
dimensions of an accretion disk that emits low ionization
lines are of the order of several thousands R$_g$. For that reason we did not consider
dimensions of the disk larger than 100000 R$_g$. This was an important 
approximation to limit the computing time.

v)  We consider a systemic velocity shift of the { non-disk} region  (not greater than
$\pm$ 3000 km/s) to test the possibility of the outflow/inflow.

\subsection{Flux ratio, normalized widths and asymmetry parameters}

We consider the following parameters to study the simulated and observed BELs profiles:

i) The flux ratio between the disk ($F_{d}$) and the
non-disk region ($F_s$):

$$Q={F_{s}\over F_{d}}\eqno(4),$$

{ where}

$$F_{tot}={F_{s}+F_{d}}=(1+Q)F_d$$

  Using this parameter, { the total  line profile (normalized to the  disk flux)\footnote{ This is taken from technical reasons to simulate different contributions of the disk and the non-disk component. First we normalized both line profiles to their fluxes, and after that we rescaled the non-disk component multiplying with Q, then whole profile is given in units of the disk flux.}} can be written as: 

$${I_{tot}(\lambda)\over F_d}={I_{d}(\lambda)\over
F_{d}}+Q{I_{s}(\lambda)\over F_{s}},\eqno(5)$$  where  $I(\lambda)$ is the wavelength dependent
intensity. The composite profile is normalized according to, $$\Im(\lambda)={I_{tot}(\lambda)\over
I_{tot}^{max}},\eqno(6)$$ where $I^{max}_{tot}$ is the maximum intensity of the composite line profile.

ii) For the composite line profile $\Im(\lambda)$ we measured full widths at 10\%, 20\%, 30\% and 50\%
of the maximum intensity, i.e. $w_{10\%}, \ w_{20\%},\  w_{30\%}$ and $w_{50\%}$. Then we define
coefficients $k_i$ ($i=10,\ 20,\ 30$) normalized to the Full Width at Half Maximum (FWHM), as
$k_{10}=w_{10\%}/w_{50\%},\ k_{20}=w_{20\%}/w_{50\%}$ and $k_{30}=w_{30\%}/w_{50\%}$. It is obvious
that the coefficients $k_i$ are functions of the radius $R$ and other parameters of the disk. Using
these normalized widths we can compare results from AGN with different random velocities.

iii) We also measured the asymmetry ($A_i$)  at $i=$10\%, 20\%, 30\% of  maximum  intensity   of the
modeled and  observed lines as

$$A_i={{W_i^R-W_i^B}\over{FWHM}},\eqno(7)$$ where $W_i^R$ and $W_i^B$ are red and  blue half widths at
$i=$ 10\%,  20\%  and 30\% of the maximum intensity, respectively.

\subsection{Simulated line profiles}

First of all, we simulated only the disk profiles, taking into account different values of   the disk
parameters. An extensive discussion about possible disk line profiles is given in  \citet{Dum90}.   In the first instance, the
relative importance of the disk  contribution  to the core or to the wings  depends on the  disk
inclination. In  { Fig. 3}  we  presented simulated profiles, corresponding to
$R_{inn}=400\  \rm Rg$,  $R_{out}=$ 1200 Rg and 12000 Rg, for  different inclinations: $i=$ 1$^\circ$,
10$^\circ$, 20$^\circ$, 40$^\circ$ and  60$^\circ$. As it can be seen in Fig. 3,  the contribution of
the disk to the center of the line or to the wings  is not so much sensitive to the outer radius, but 
significantly depends on  the disk inclination. A face-on disk contributes more to the
core of the line, while a moderately inclined disk ($40^\circ >i>20^\circ$) contributes significantly to the line wings.  For $i>40^{\circ}$  the  disk emission will strongly affect  the far wings of the composite profile.

\begin{figure*}
\centering
{\includegraphics [height=6cm,width=5cm,angle=-90] {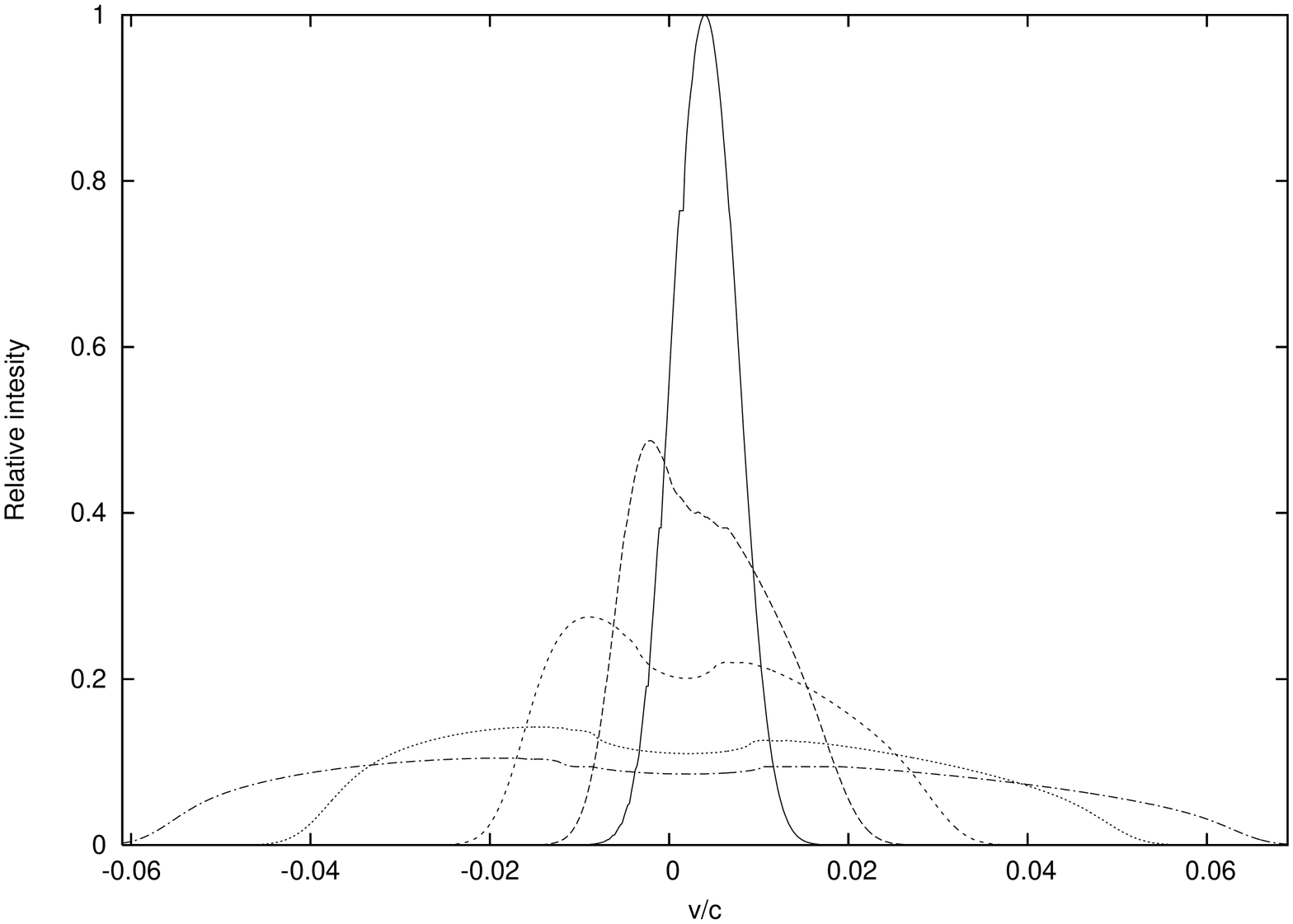}}
{\includegraphics [height=6cm,width=5cm,angle=-90] {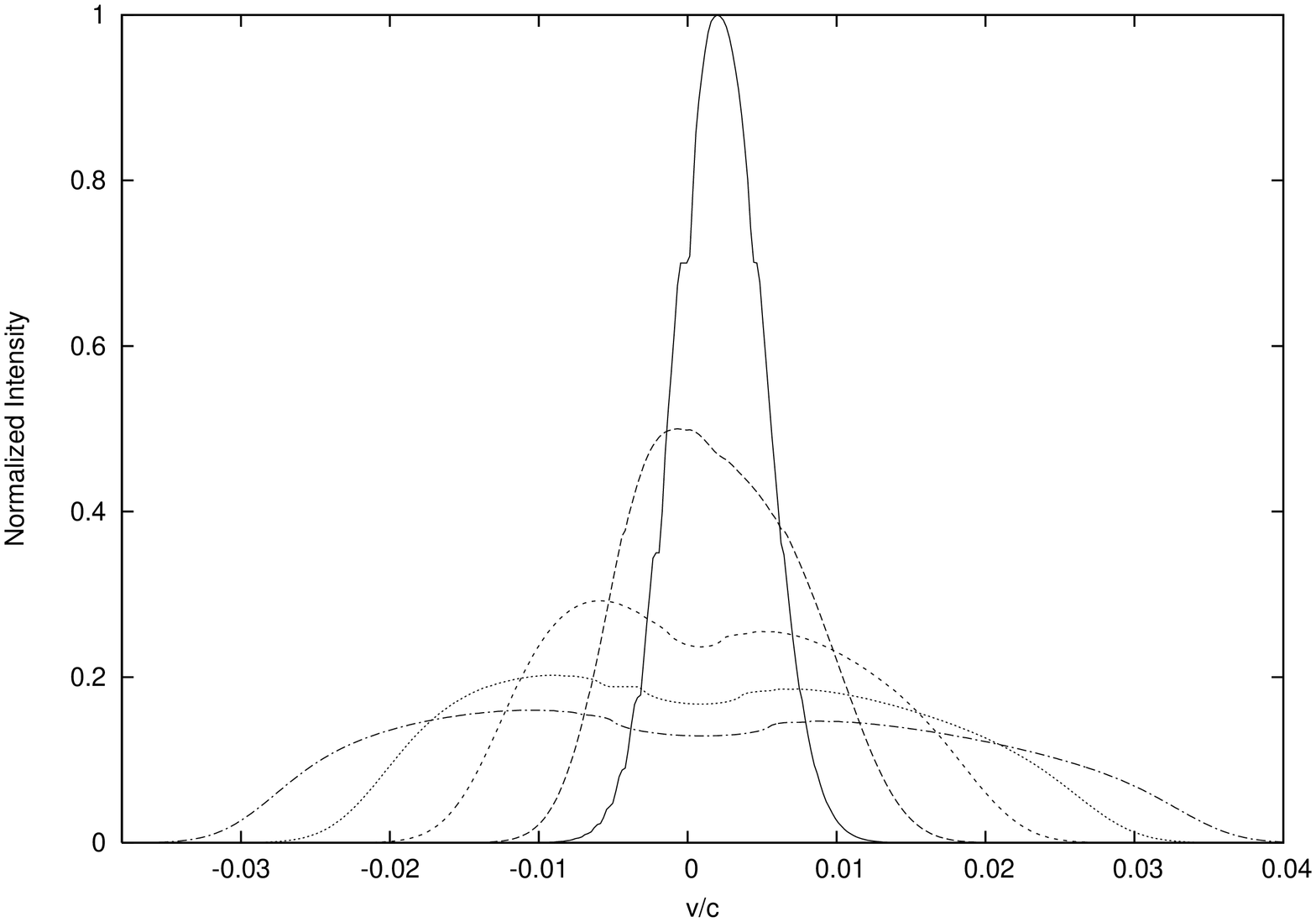}}
 \caption{Simulated disk line profiles for five different disk inclinations ($i$=1, 
10, 20, 40 and 60 degrees, from the highest to the lowest intensity lines, 
respectively) for the disk with the fixed inner
  radius $R_{inn}=$400 Rg and with outer $R_{out}$=1200 Rg (left) and   
$R_{out}=${12000} Rg.(right).} \label{fig:Simulation}
\end{figure*}

Another very important parameter is the flux ratio between components,  $Q$. As examples, in Fig. 4 we
presented five  simulations of composite line profiles with values $Q=0.3,\ 0.5,\ 0.7,\ 1$, 1.5 and 2,
where  $R_{inn}=400  \rm \ Rg$  and $R_{out}=1200\rm \ Rg$, and for different inclinations ($i=$1, 10,
20,  40, 60 degrees).

\begin{figure*}
\centering
\includegraphics [height=5cm,width=4cm,angle=-90] {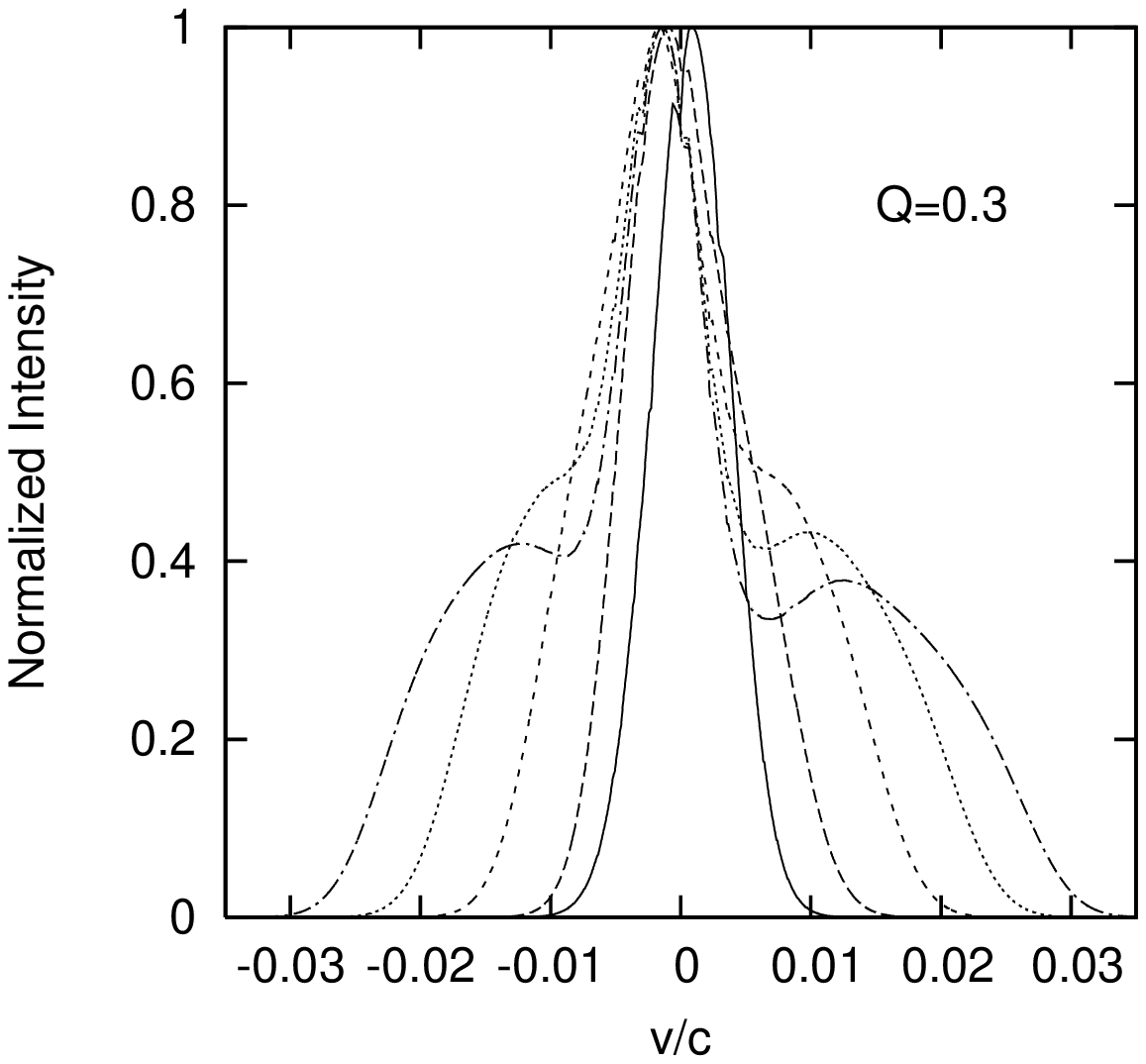}
\includegraphics [height=5cm,width=4cm,angle=-90] {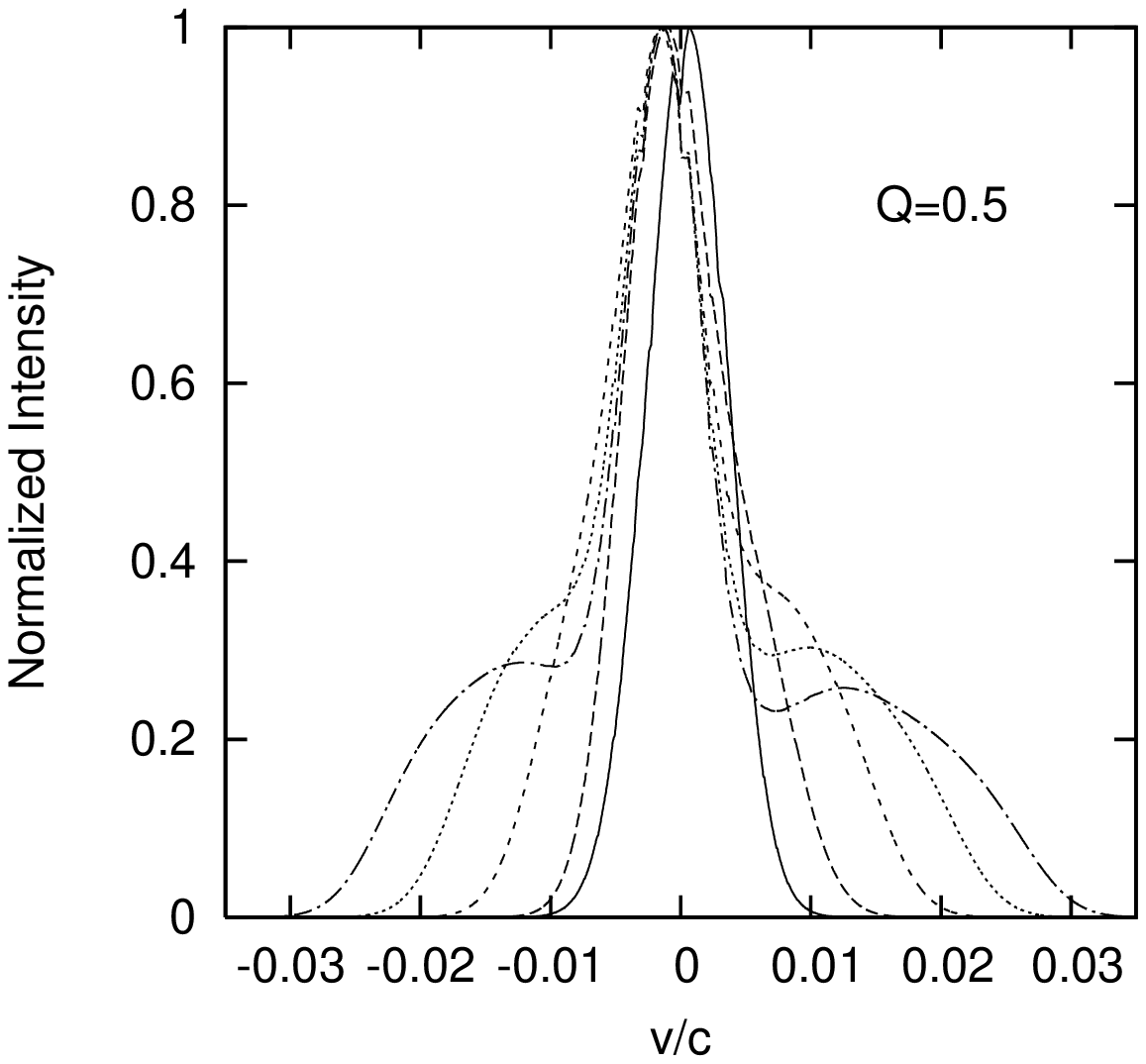}
\includegraphics [height=5cm,width=4cm,angle=-90] {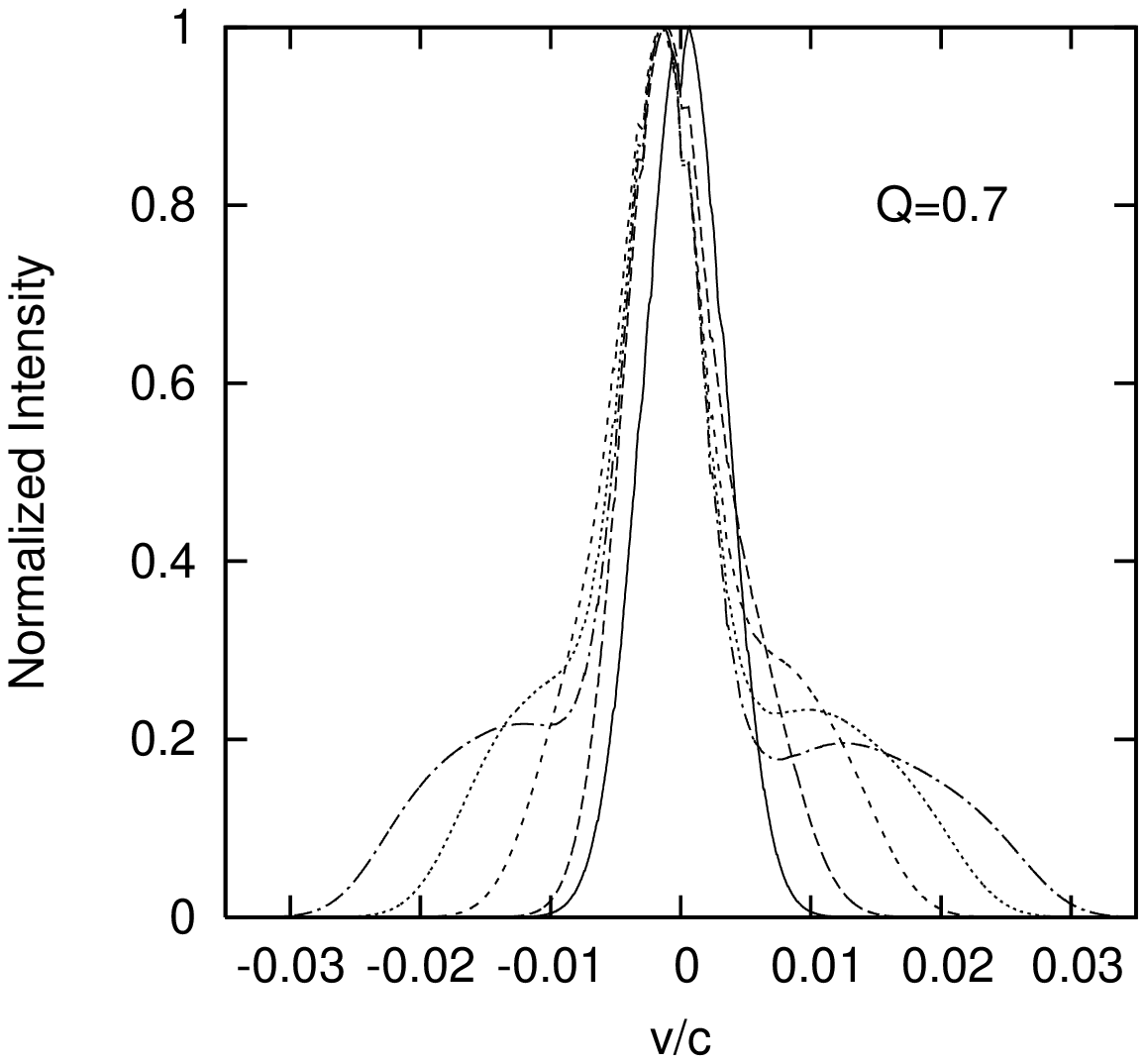}
\includegraphics [height=5cm,width=4cm,angle=-90] {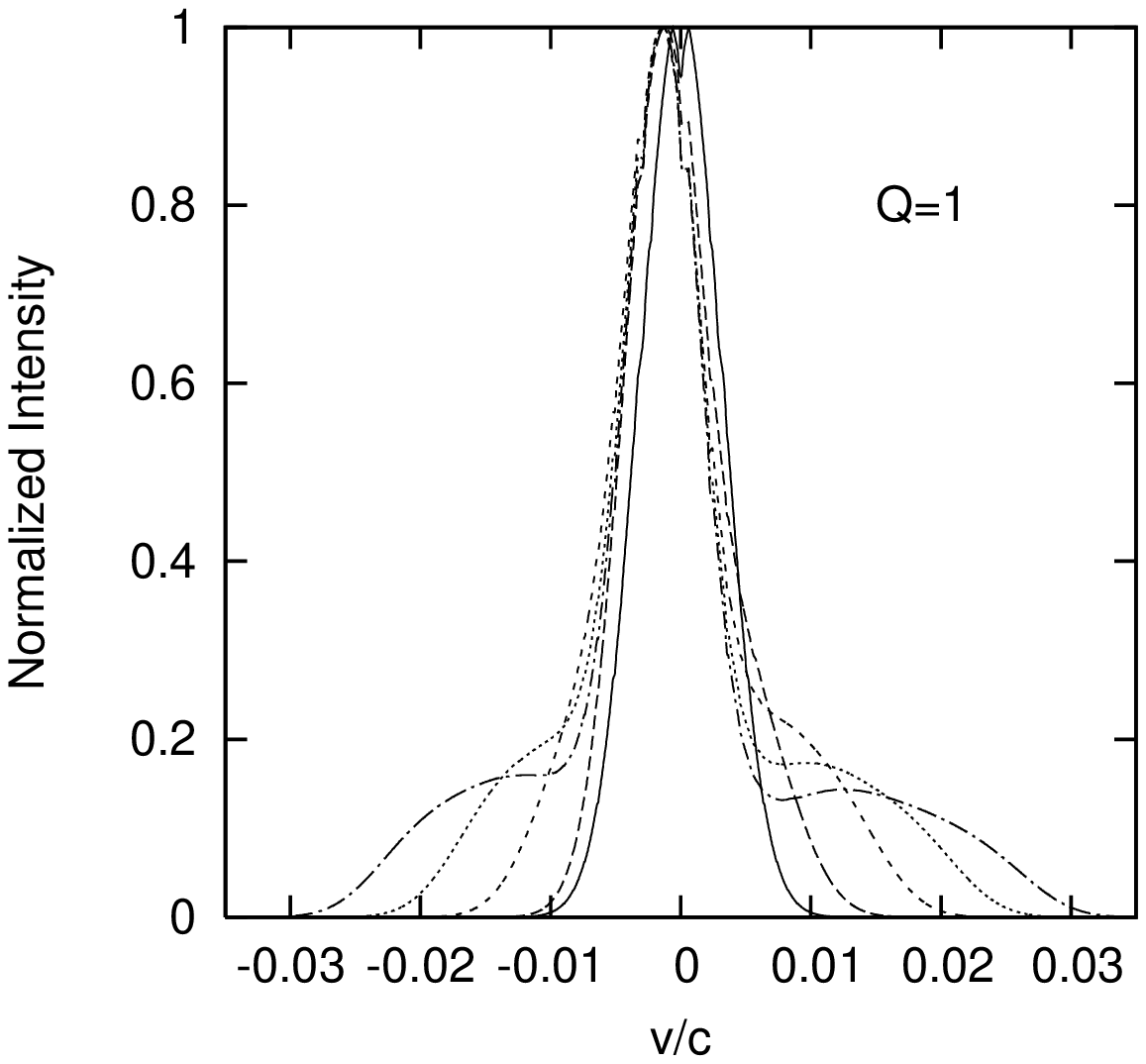}
\includegraphics [height=5cm,width=4cm,angle=-90] {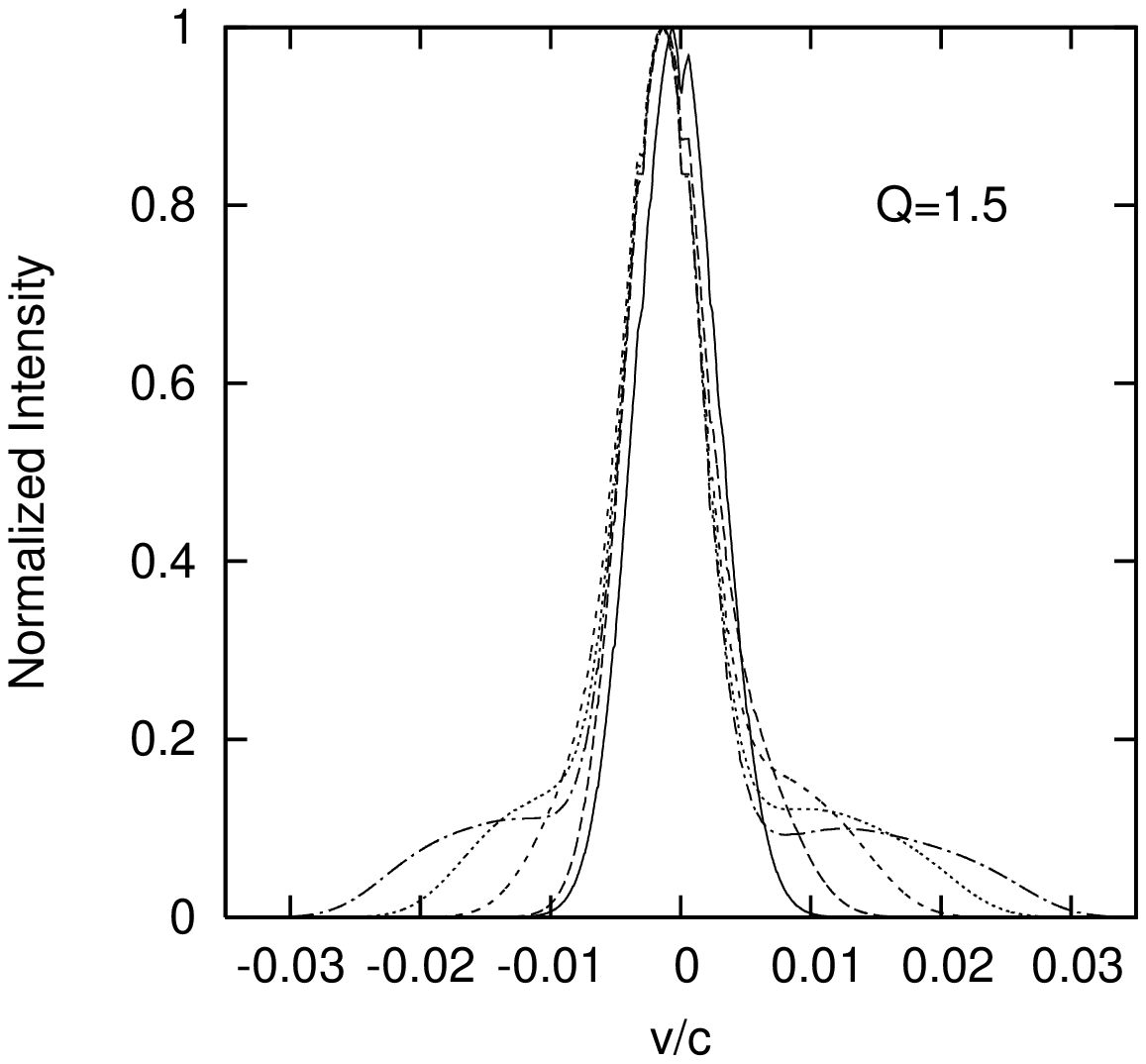}
\includegraphics [height=5cm,width=4cm,angle=-90] {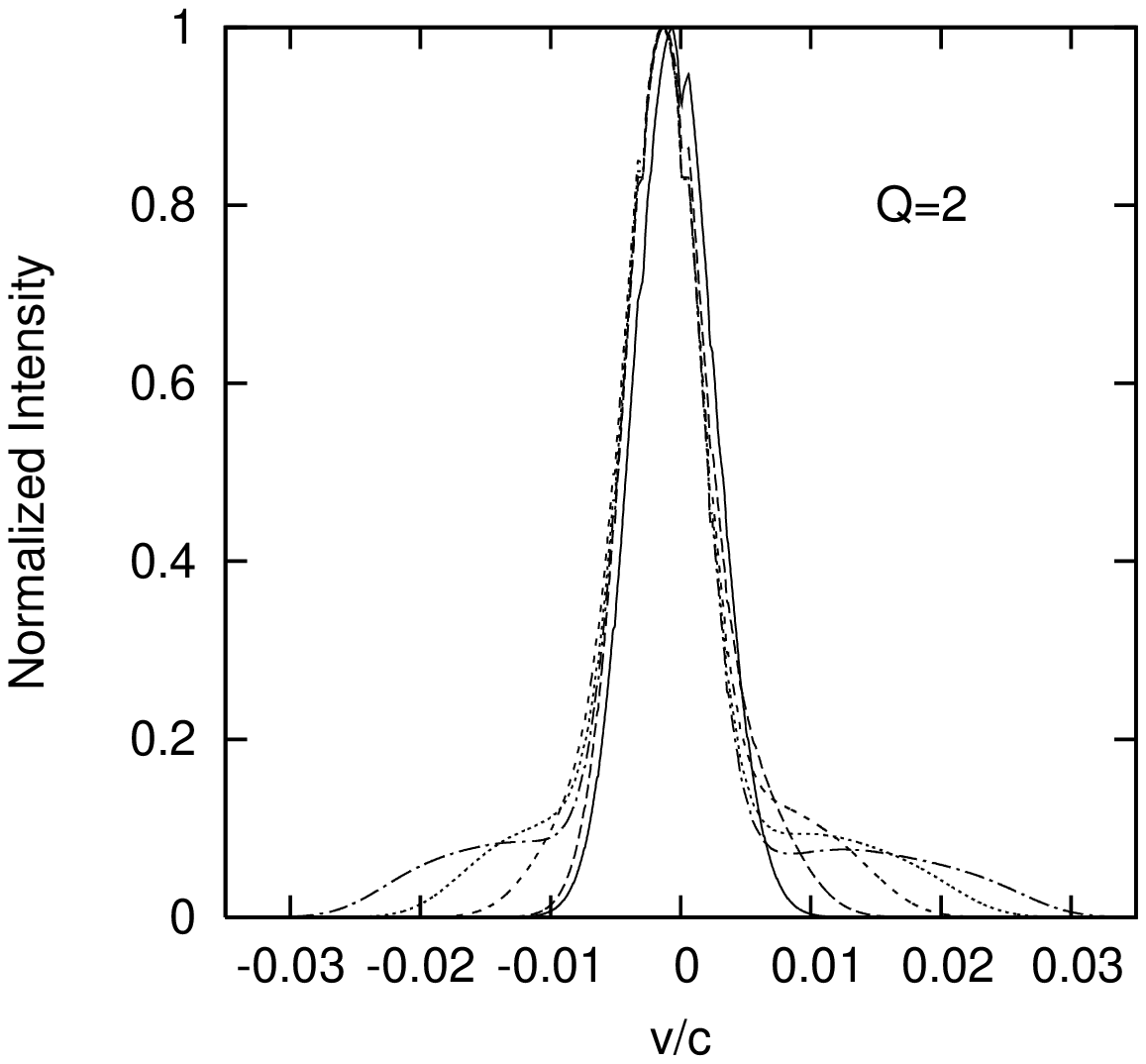}
 \caption{Simulated line profiles emitted by the two-component model for 
five different inclinations ($i$=1,
10, 20, 40 and 60 degrees, from the narrowest  to the broadest 
line, respectively) for different contributions of the disk to the composite line 
profiles (as it is written in figures). The inner radius of the disk 
is taken to be 400 Rg, while the outer radius is 1200 Rg.
}\label{fig:Simulation1}
\end{figure*}

We found that  the presence of the disk 
emission is difficult to detect in the line profile when the contribution of the disk  is
smaller than 30\% of the total line emission ($Q>2$): in the case of a low inclination both the disk and non-disk region contributes to the line core and it is very hard to separate the disk and nod-disk region. 
In the case of a highly inclined disk,  the disk emission spreads
in the far wings, and could not be resolved from the continuum, especially if the observed spectrum is noisy. 
For the case of dominant disk emission ($Q<0.3$), 
if the inclination is low, the line will be shifted  to the red, and if the inclination 
is high,  two  peaks or at least shoulders should appear in the 
composite  line profile. 
Consequently, 
further in the paper we will consider only cases where  0.3$<Q<$2.  

Note 
here that, although we
have considered a relatively low random velocity of $\sim$1000 km/s, the lines where the disk is dominant (the disk contributes at least 50\%) can be very broad. The obtained widths are in agreement with the measured widths 
of double-peaked lines, that range from several 
thousand \citep{EH94,EH03} to nearly 40,000 km s$^{-1}$
 \citep[see e.g.][SDSS J0942+0900 has the H$\alpha$
width of $\approx$ 40000 km s$^{-1}$]{wang05}.

\subsection{Results from the line profiles simulations}

From the  simulations { mentioned above} we infer the following results:

(i) To detect disk emission in a BELs, the fraction of the flux 
emitted by the disk in the total line profile should be higher than 30\%. A dominant  disk ($Q<0.3$) will be 
clearly present in the total line profile (peaks or shoulders in the line profiles). 

(ii) In the case of a nearly face-on disk ($i<5^\circ$), the disk emission may contribute to a
slight asymmetry towards the red (due to the gravitational redshift), 
but it is hard to detect this asymmetry. 
In the case of an edge-on disk, the emission from the disk will contribute to the far wings and then it may be difficult to separate it from the continuum.

(iii)  These two parameters, the flux ratio between components and disk inclination, are crucial for the line shapes in the two-component model.

(iv) In the simulated line profiles the asymmetry 
was mostly $A_i>0$. For low inclination ($i<10^\circ$), the asymmetry weakly depends on 
$Q$.

\section{Comparison between simulated and observed BEL profiles}

\subsection{Data sample and measurements}

{ The set of spectra for our data sample has been collected by \citet{Gio07} 
from the 
 spectral database of the third data release from the Sloan Digital Sky 
 Survey (SDSS)\footnote{http://www.sdss.org/dr3}. 
 According to the purposes of the work  \citep[see also][]{Gio07}), the SDSS database 
was searched for 
 sources corresponding to the following requirements: 
(i)  objects were with redshifts $z<0.4$, so  H$\alpha$ would be 
covered by the available spectral range; 
(ii) the Balmer series were clearly recognized, at least up to H$\delta$ in order to see that in all
Balmer lines a broad component is present; 
(iii) and profiles were not  affected by distortions (bad pixels on the 
sensors, the presence of strong foreground or background sources).

 The preview spectra provided by the database retrieval software were 
 manually inspected, looking for objects in better agreement with our 
 requirements, until 115 sources were chosen from approximately 600 
 candidates examined in various survey areas. Subsequent inspection of the 
 spectra collected within the database led to the rejection of 25 objects, 
 which were affected by problems that could not be detected in the preview 
 analysis. Therefore, our resulting sample includes the spectra of 90 
 various broad-line-emitting AGN, corresponding to 15\% of the candidates 
that we examined and located in the range.

 The spectra were already corrected for instrumental and environmental 
 effects, including sky-emission subtraction and correction for telluric 
 absorption and calibration of data in physical units of the flux and 
wavelength. 
 Spectra were corrected for the Galactic extinction \citep[see ][]{Gio07}. 
Also, the  cosmological 
redshift calibration were performed.
 Since the interest was to investigate the broad line shapes, the 
 subtraction of the narrow components of H$\alpha$ as well as the 
satellite [NII] 
lines were performed. 
The spectral reduction (including subtraction of stellar component) and the way to obtain
the  broad line profiles are in more details explained in \citet{Gio07}. 
We have also used FWHM and FWZI measurements  from Table 2. of the 
mentioned paper.

\begin{figure}
\centering
\includegraphics[width=8.5 cm]{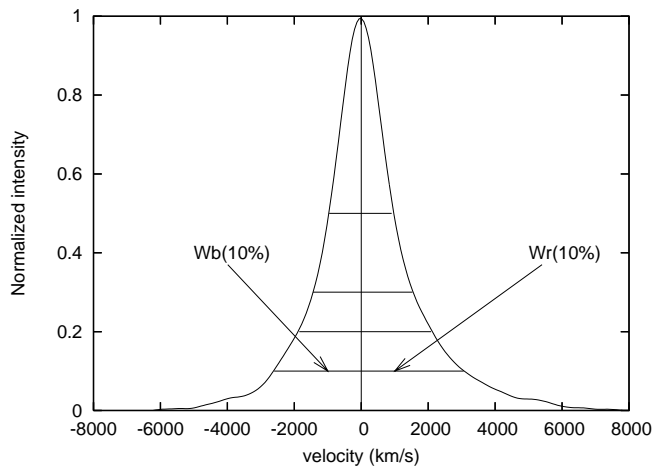}
\caption{The broad H$\alpha$  line of SDSSJ1025+5140.
 The horizontal lines presented the measured
widths at 10\%, 20\%, 30\% and 50\% of the maximum intensity. The blue and red half
widths have been measured.}
\end{figure}

Previously cleaned broad H$\alpha$ profile was 
normalized, converted from wavelength to velocity scale and smoothed, with 
Gaussian smoothing, using DIPSO software package\footnote{http://www.starlink.rl.ac.uk}. Additionally, we have measured the half  
(red and blue) widths, $W_i^R$ and $W_i^B$, at 10\%, 20\%  30\%  and 50\% of 
maximum intensity (see Fig. 5), and after that we calculated normalized 
full widths, $k_i$, and asymmetries, $A_i$ (Eq. (7)).
}

\subsection{Observed vs. simulated line profiles parameters}

In Fig. 6 we presented the normalized widths $k_{20}$ vs. $k_{10}$ and $k_{30}$ vs.  $k_{10}$ measured for  H$\alpha$ lines of the
sample. We  also plot the values corresponding to simulations with different disk inclinations, flux ratios $Q$, and fixed disk dimensions of $R_{inn}=600$ Rg and ${R_{out}=4000} \ {\rm Rg}$.  As one can see  in Fig. 6, most of the measured points 
are located within $0.7<Q<2$, and $i<25^\circ$. These results do not change significantly if we change the inner and outer
radii of the disk, where the inner radius is not taken to be less then 100 $Rg$, while the outer was not taken to be closer then several hundreds $Rg$.

\begin{figure*}
\centering
\includegraphics[width=8 cm, angle=270]{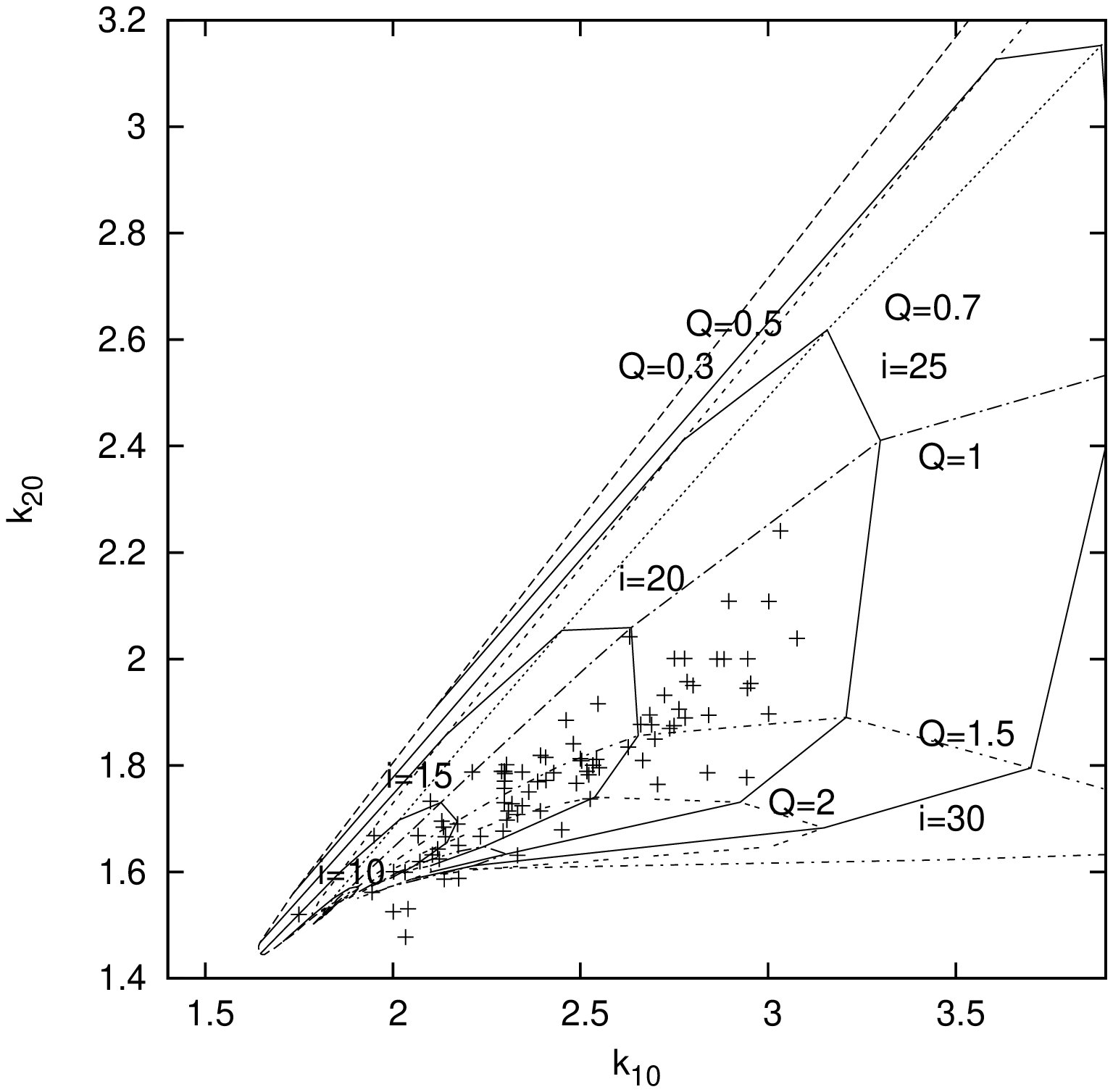}
\includegraphics[width=8 cm, angle=270]{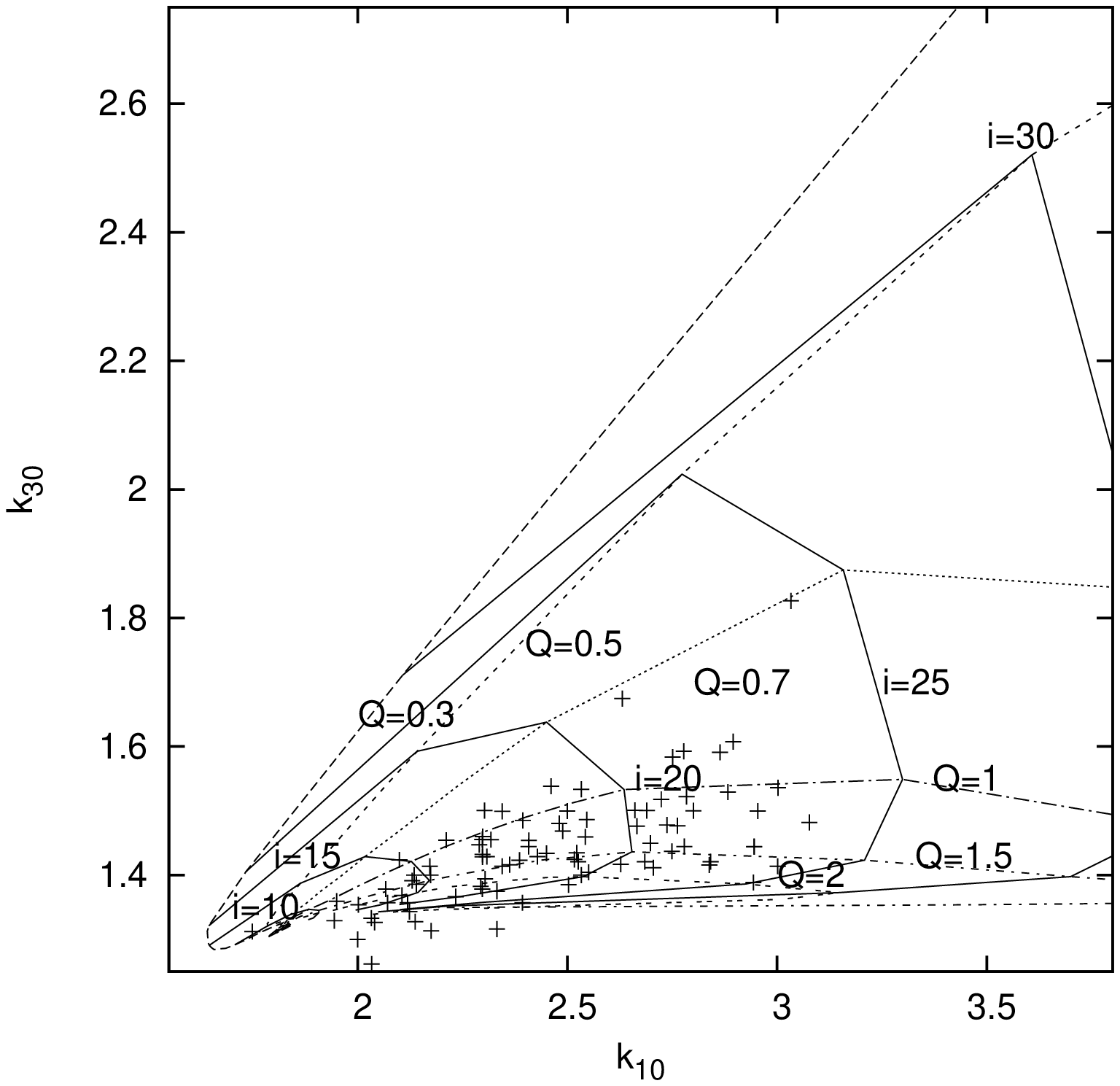}
\caption{The measured width ratios  (crosses) and simulated 
values (dashed lines) from the two-component model for the different  
contribution of the disk 
emission to the total line flux ($Q$=0.3, 0.5, 0.7, 1, 1.5 { and 2}). The  
inner disk radius is taken to be 600 Rg, outer 4000 Rg, and different 
inclinations are considered (solid  isolines  presented   {{$i$ = 10, 15,
20, 25 and 30}} degrees, respectively). 
}
\end{figure*}

The measured asymmetries of the H$\alpha$ profiles are presented in Fig. 7. Lines  in this Figure correspond to the
simulated models as in Fig. 6.  As it can be seen in Fig. 7,
some objects  (around 10 objects from the  sample of  measured H$\alpha$
lines) have a high negative  asymmetry which cannot be explained by assumed models
without a systemic motion of the non-disk region. To explore the outflow or inflow presence (to explain the blue asymmetry of the core component), we consider shifts of the Gaussian component in the model. As it can be seen in Fig. 7 (bottom) models where a blueshift  of the Gaussian component of  $800 \ {\rm km} \ s^{-1}$ was taken into account {(other parameters are not changed)}, are able to explain the measured asymmetries.

\begin{figure*}
\centering
\includegraphics[width=6.5 cm, angle=270]{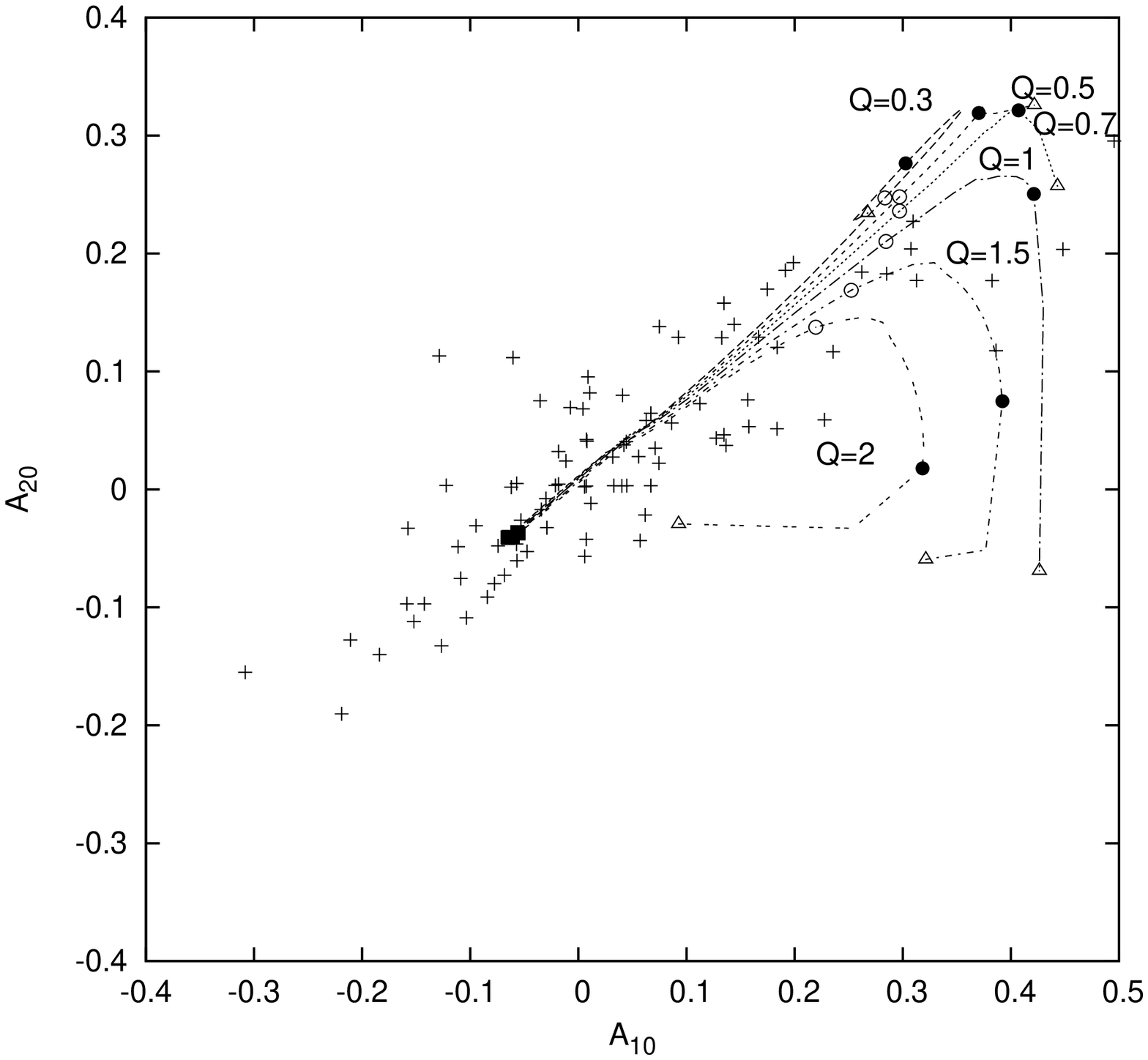}
\includegraphics[width=6.5 cm, angle=270]{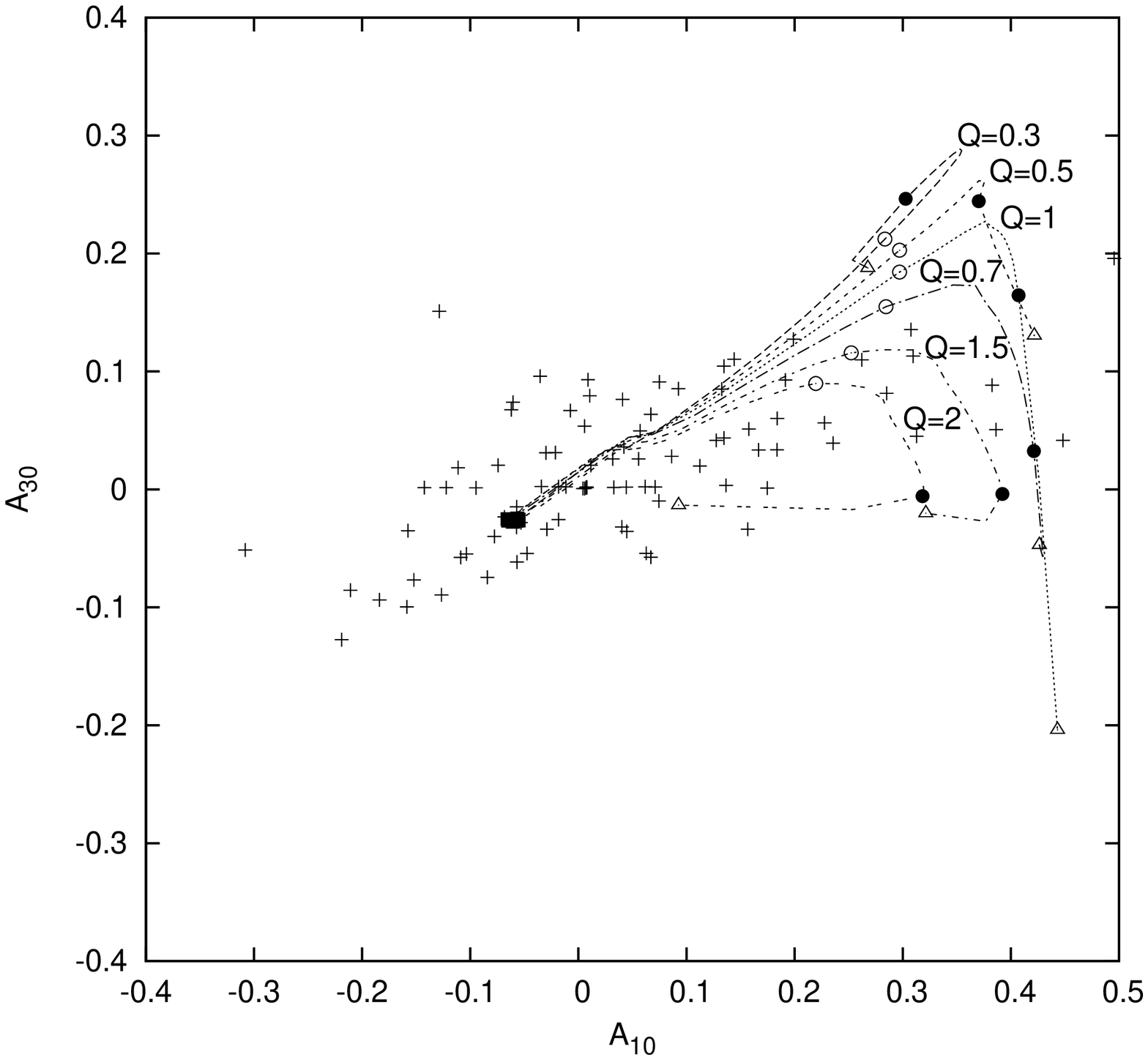}
\includegraphics[width=6.5 cm, angle=270]{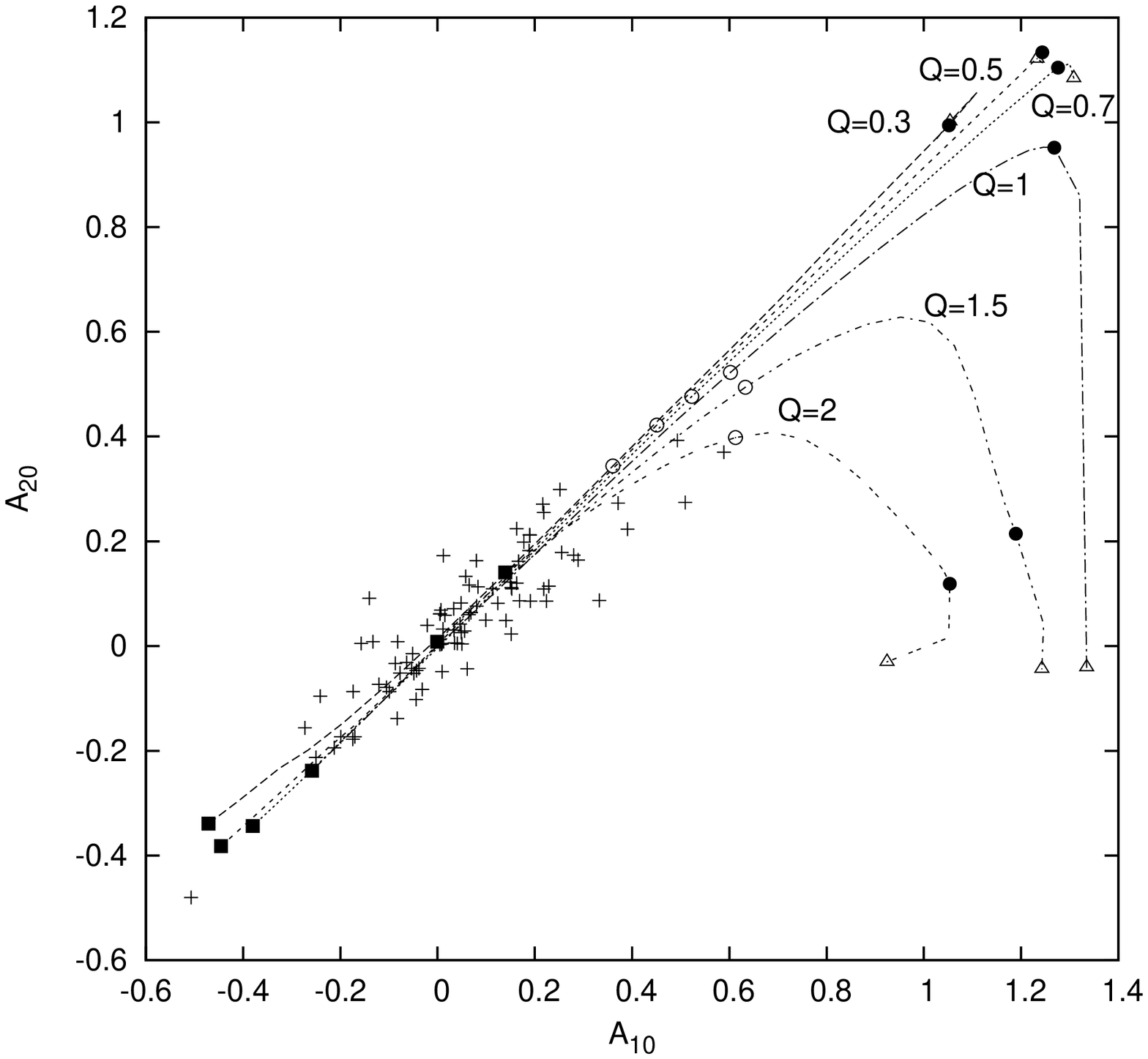}
\includegraphics[width=6.5 cm, angle=270]{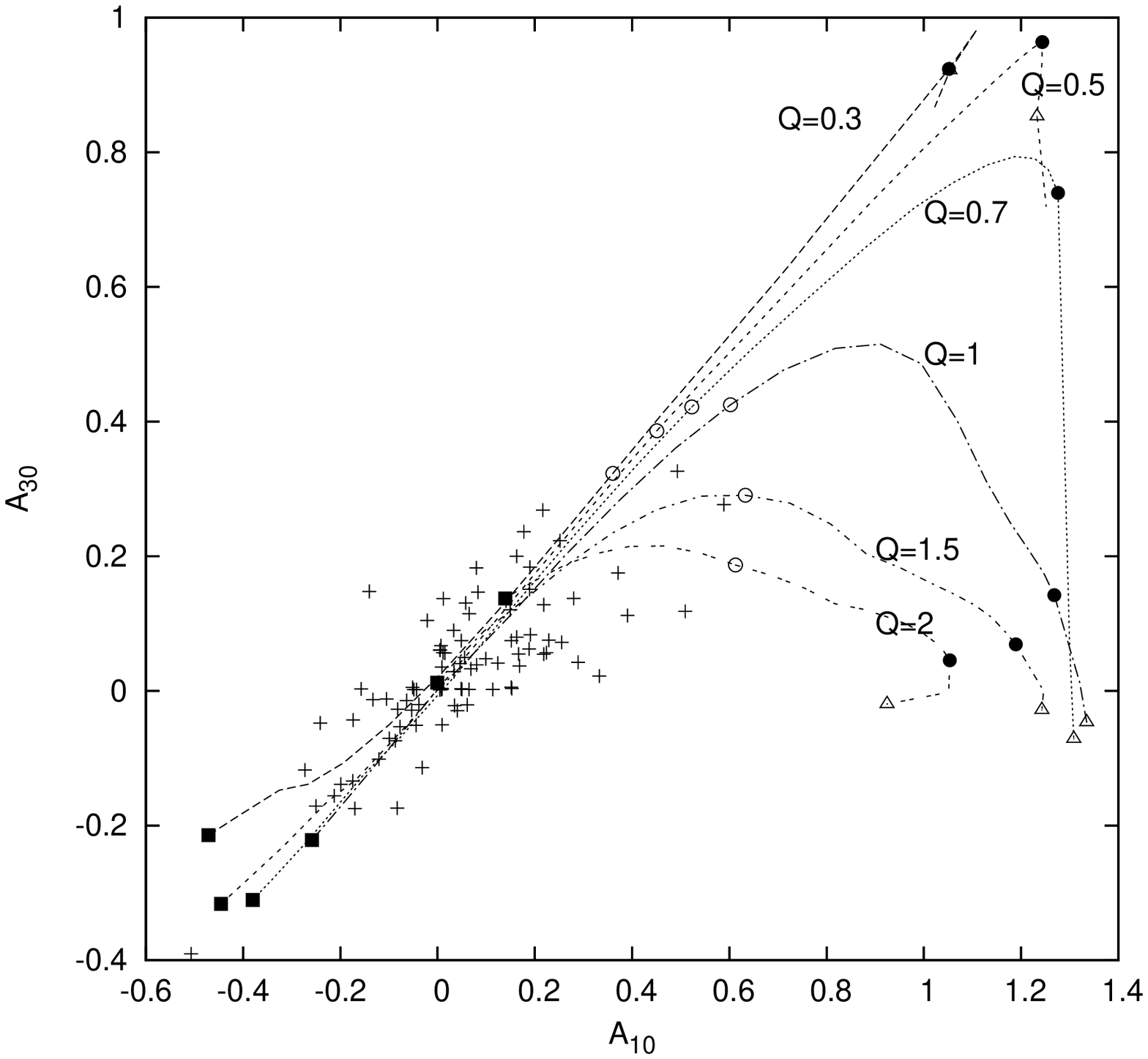}
\caption{The asymmetry of the H$\alpha$ line of the sample (crosses) and simulated
one from the two-component model. Different contributions of the disk
emission to the total line flux ($Q$) are taken into account, { and plotted with a dashed lines ($Q$=0.3, 0.5, 0.7, 1, 1.5 and 2)}. The inner disk radius is taken to be { 600 Rg and outer 4000 Rg}. Different
inclinations are considered, taking values from 0 { (full squares), 10 (open circles),   20 (full circles) and 30 (open triangles)}  degrees. { On two panels up we presented the models without systematic shift of the non disk component, while on two panels bottom, a systematic blue shift of  800 km/s was taken into account.} }
\end{figure*}

As it can be seen in Fig. 6, the coefficients $k$ are sensitive to the disk
inclination and $Q$, therefore we will use the relationships between these
coefficients to give some estimates of the inclination and $Q$ for the
sample.  Of course, we should also take into account the influence of the disk
dimensions.  To do that we compute a { grid} of models with $k_i=f(R_{inn},i,Q)$
confirming that the changes in the parameters mainly depend on $i$ and $Q$.  We
found that  for inner radius R$_{inn}<$200 R$g$, the simulated $k_i$ do not 
fit well the observations { (most of the measured points from the sample are out of the { grid} of models, as it is presented in Fig. 6)}. The same inconsistencies are found for R$_{out}<$1000
R$g$.  

In order to derive some results about the disk parameters we assume the following constraints and procedures:

i) According to the results obtained from the fitting of the double-peaked
lines in \citet{EH03} we fixed  the inner radius at R$_{inn}$=600 R$g$, and the
outer at R$_{out}$=4000 R$g$, as the averaged values obtained from their fittings
\citep{EH03}.

\begin{figure}
\centering
\includegraphics[width=8 cm]{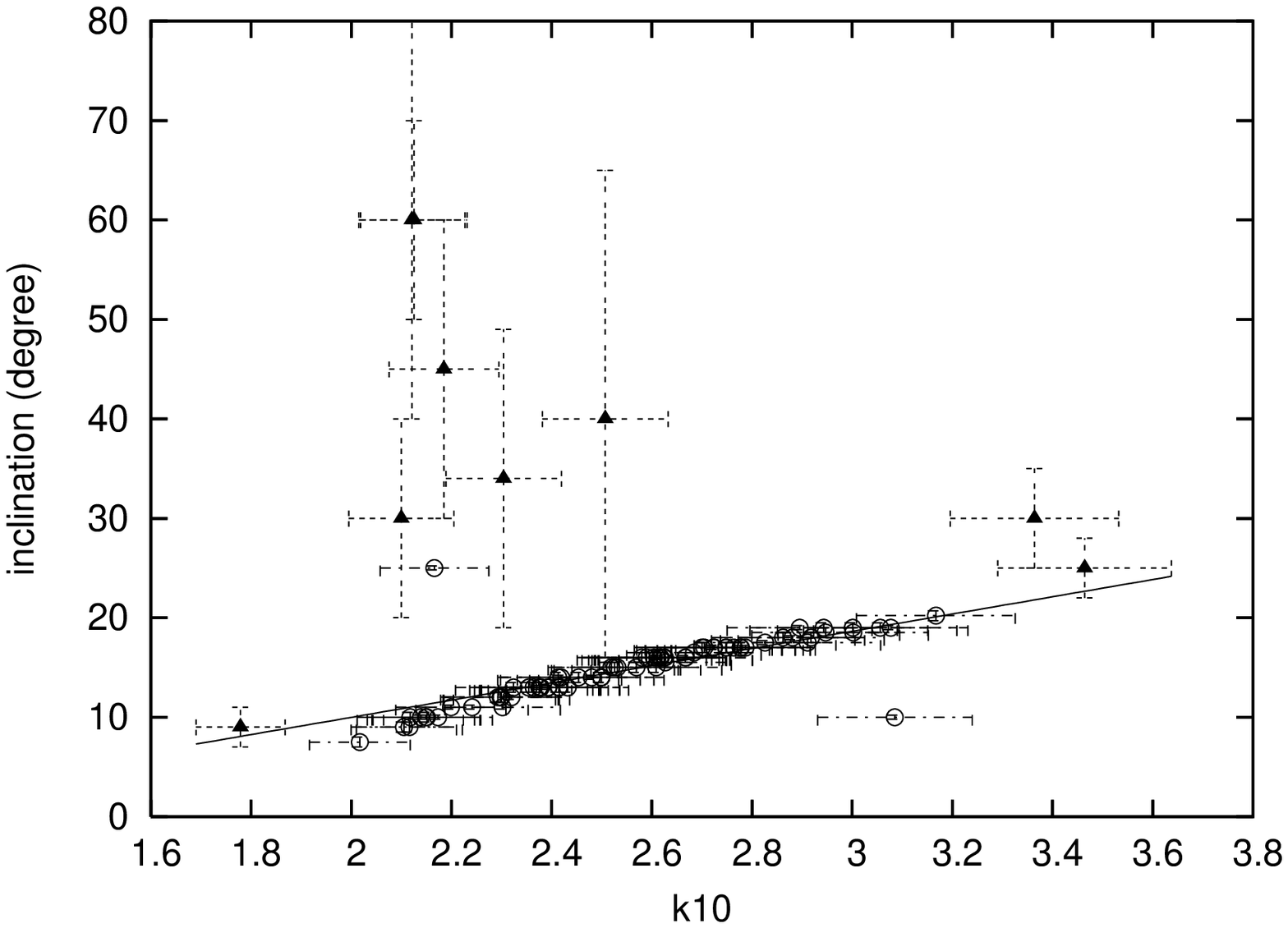}
\includegraphics[width=8 cm]{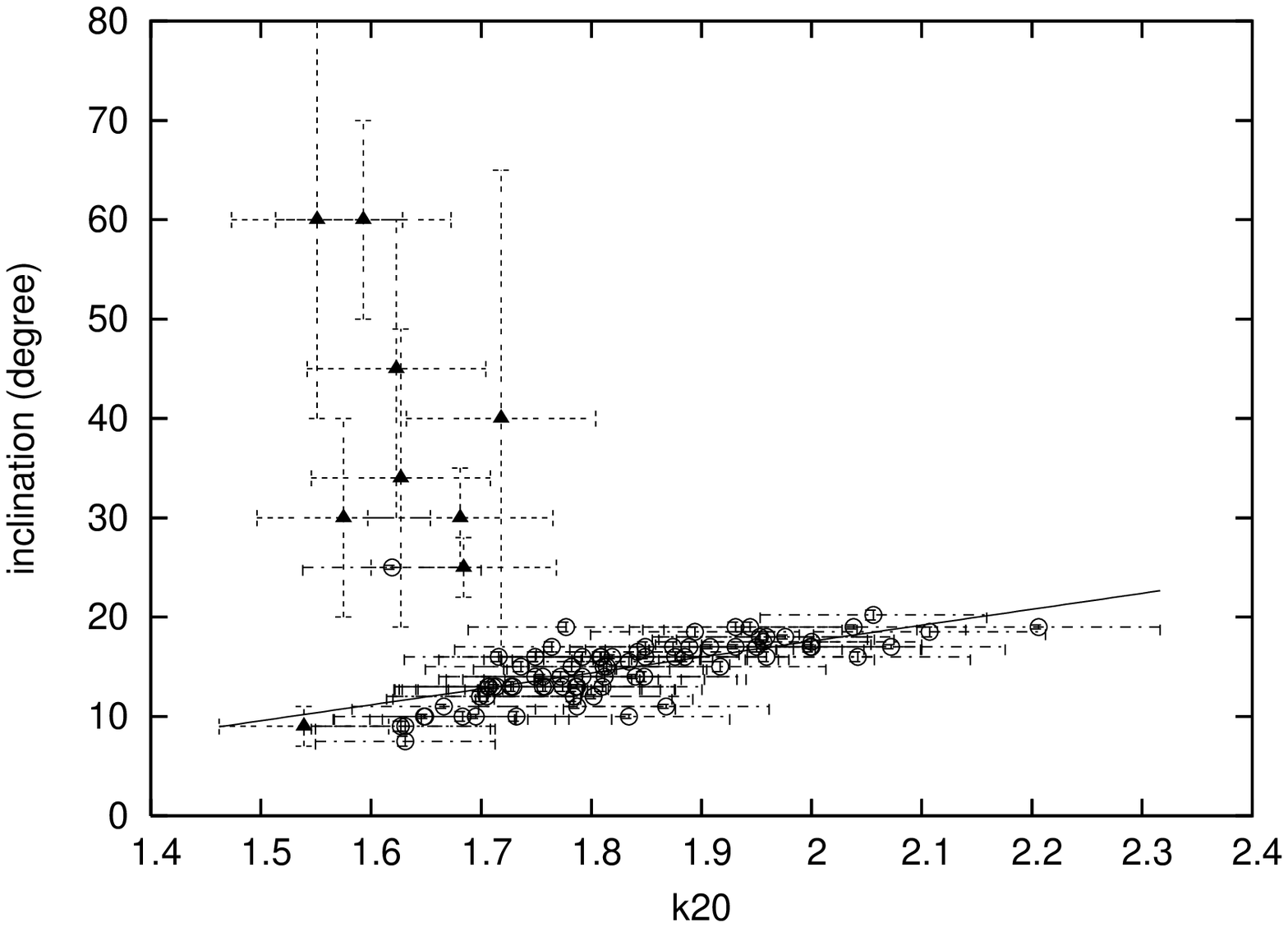}
\includegraphics[width=8 cm]{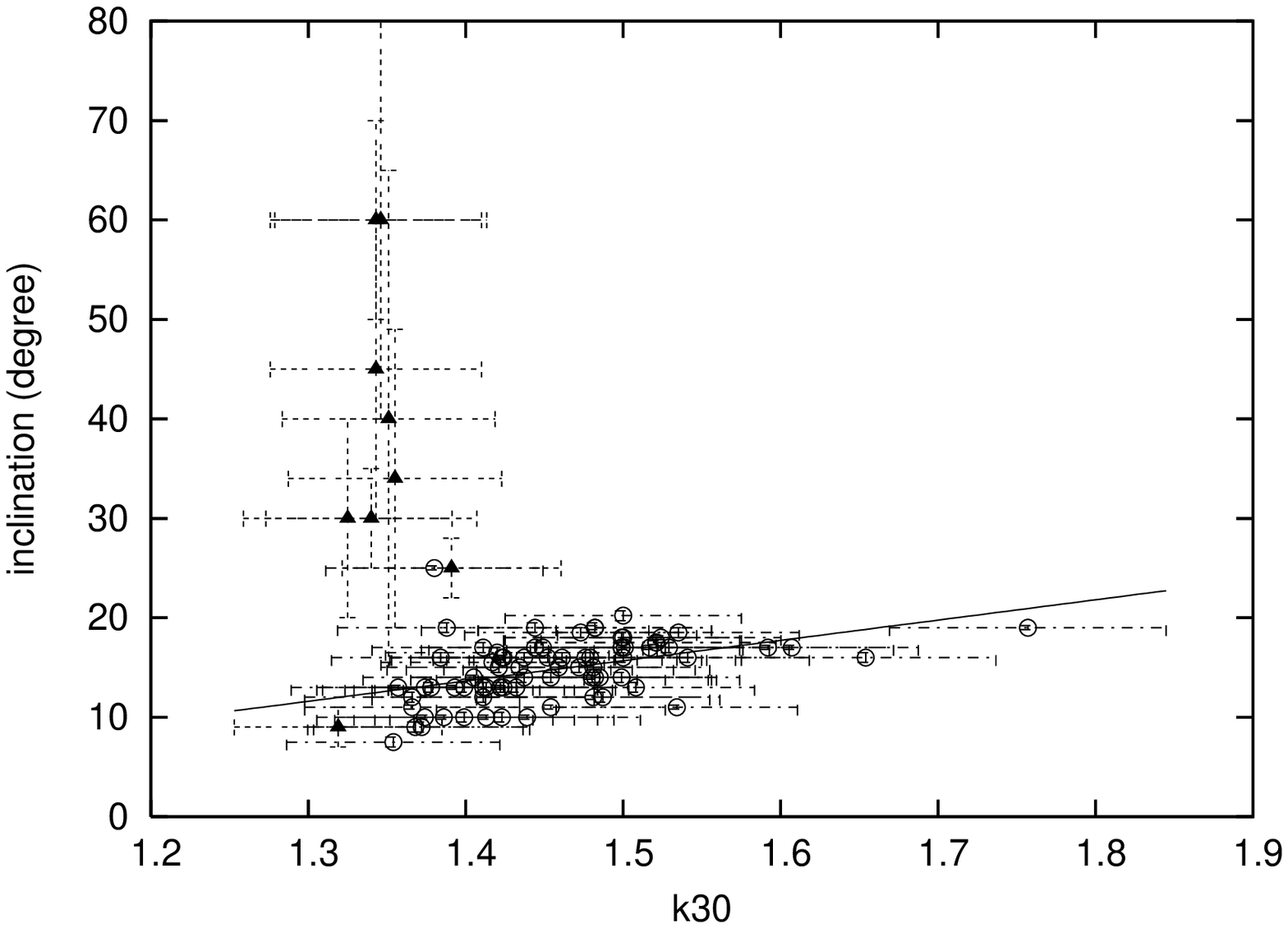}
\caption{Inclination vs. $k_{10}$ (the first),  $k_{20}$ (the second) and $k_{30}$ (the third). The points
where description were more than one degree (mainly above the other points) are denoted with 
full triangles.}
\end{figure}

ii) For each AGN in the sample we estimate $i$ and $Q$ using the normalized widths, $k_i$.
Specifically, we obtain two estimates for $i$ and $Q$ as the values associated
to the measurements of both $k_{10}$ vs. $k_{20}$ and $k_{10}$ vs.
$k_{30}$. In Table 1 we present the averaged values and differences between
those estimates, $\Delta i$, and $\Delta Q$ {for the case without blue shift of the non-disk component}.. 

iii) We excluded from the analysis the objects {where difference between estimated  $i$ and  $Q$ was huge, in total 5 objects where the two-component model cannot be applied. Also there are 9 objects $\Delta i>1\,\rm
degree$  (presented as full triangles in Fig. 8 and in Figs. after).} In Fig. 8, we present  the $k_i$ parameters as function of
the inclination. As one can see, most of the points are well concentrated as a
linear function of $i$ vs. $k_i$.

\begin{figure*}
\centering
\includegraphics[width=6 cm,angle=-90]{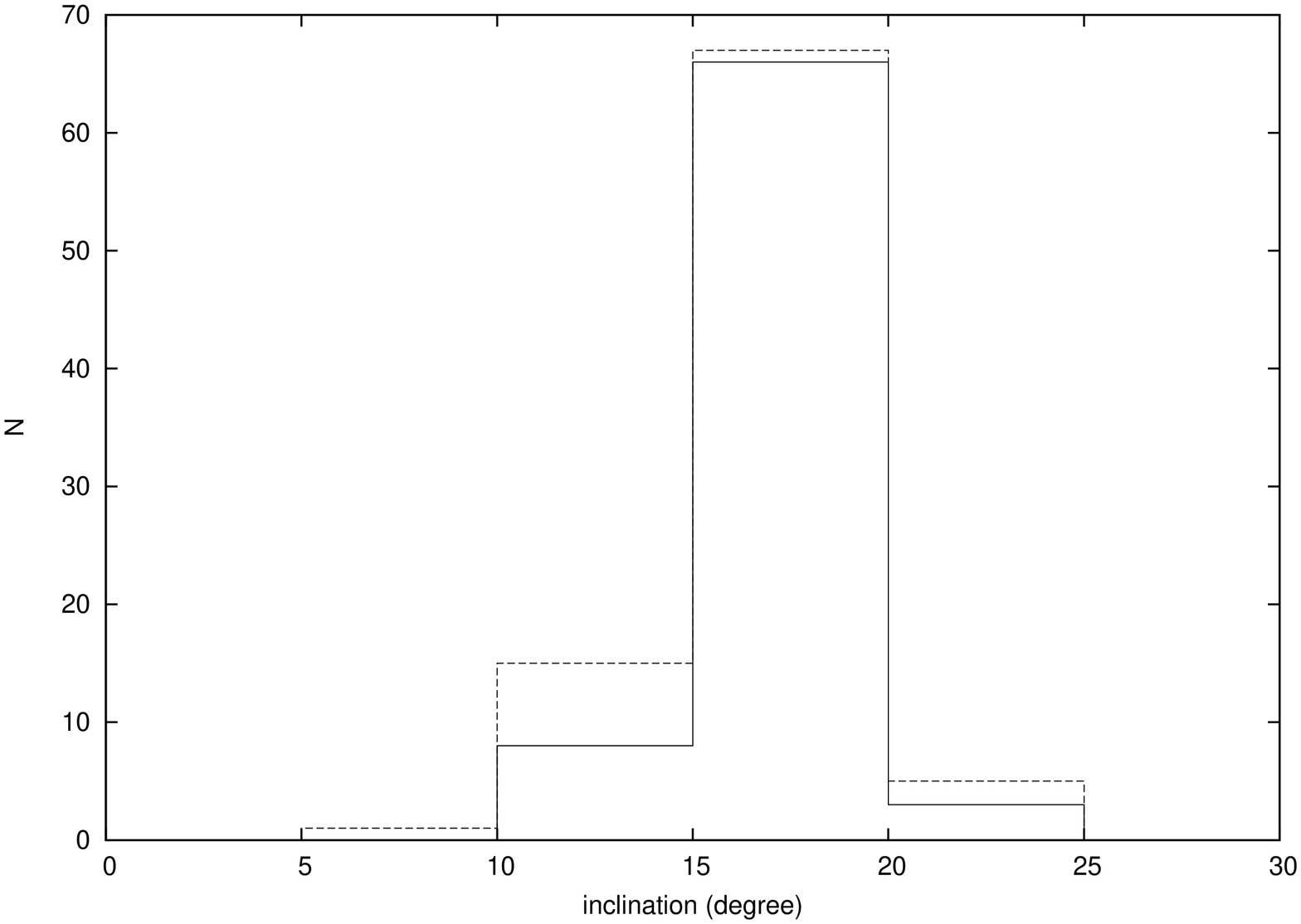}
\includegraphics[width=6 cm,angle=-90]{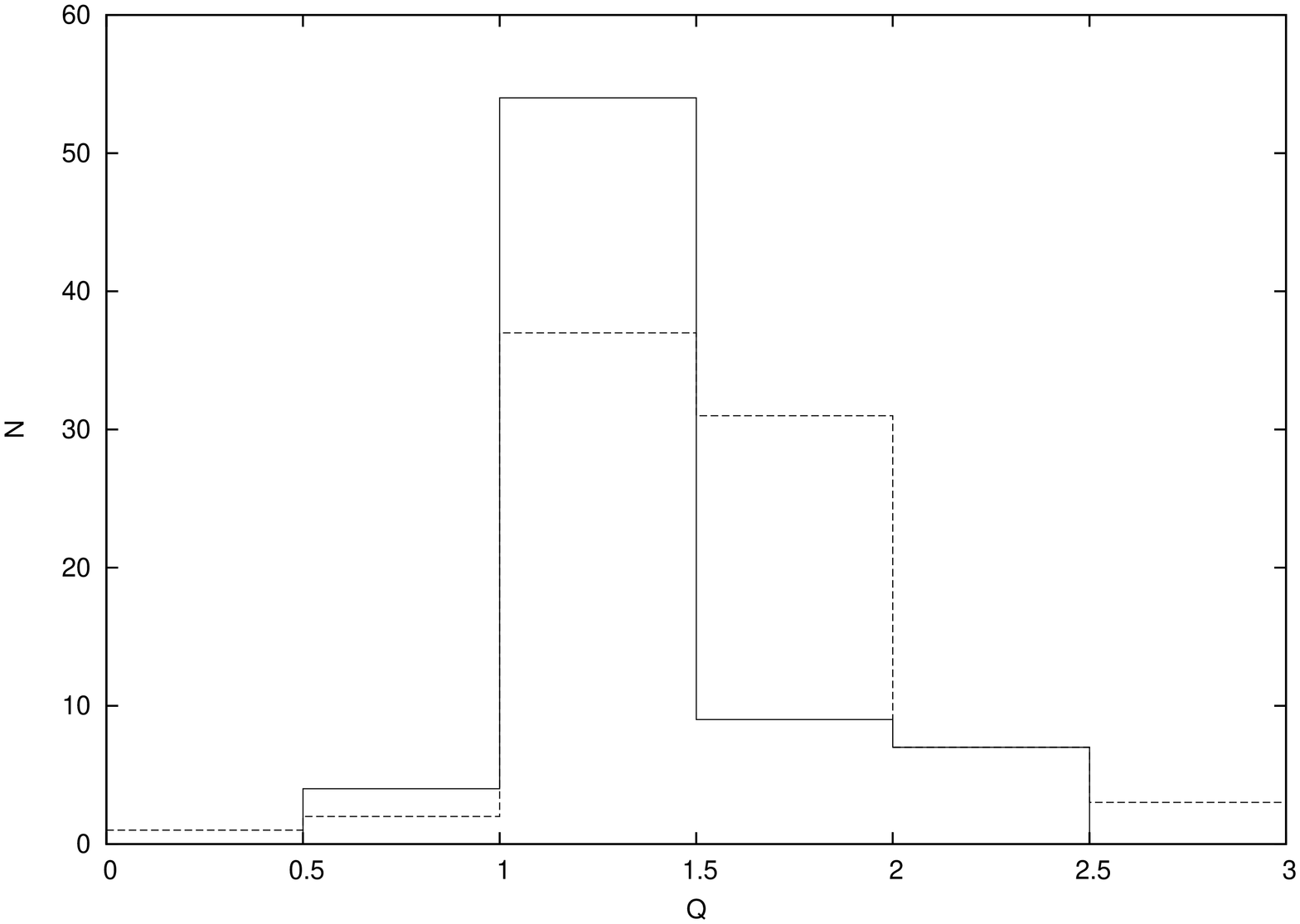}
\caption{Histograms of the inclination (left) and $Q=F_s/F_{disk}$ (right) for the sample {in the case with (dashed line) and without (solid line) systematic blue shift of the non-disk component (see Fig. 7)} . The 
cases where discrepancy in the inclination were more than one degree, were not taken into account.}
\end{figure*}

In Fig. 9, we present histograms of  the number of AGN vs. $i$ and 
$Q$. { As one can see in Figure, there is a peak at 
$i\sim17^\circ$ while estimated values are within $10^\circ<i<25^\circ$ { in both cases: without (solid line) and with  systematic blue shift of the non-disk (see Fig. 7)}. Also, there is a peak at
$Q\sim 1.25$ and most of the points are within  $1<Q<2$, showing that the disk emission is typically smaller than the emission of the non-disk region.}

One can expect randomly oriented accretion disk in AGN, but we obtained low inclined disk. Such small inclination range  ($10^\circ<i<25^\circ$), { may} be expected since a highly inclined disk { has} a smaller brightness than surrounding non-disk region. { Therefore one can expect a weak disk emission in far wings which cannot be detected.}

\begin{figure}
\centering
\includegraphics[width=4.5 cm, angle=270]{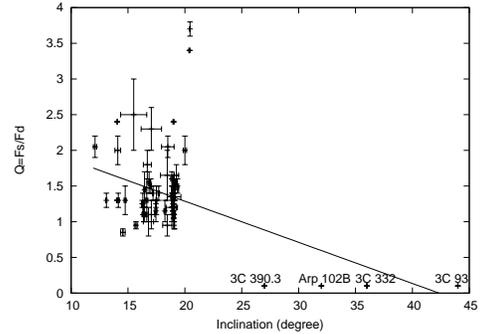}
\caption{Q vs. inclination for the sample, and the position of the well known AGN 
with double-peaked lines (Eraclous \& Halpern 1994), where is assumed that in these
objects $Q<0.1$. } \end{figure}

\begin{figure}
\centering
\includegraphics[width=7 cm]{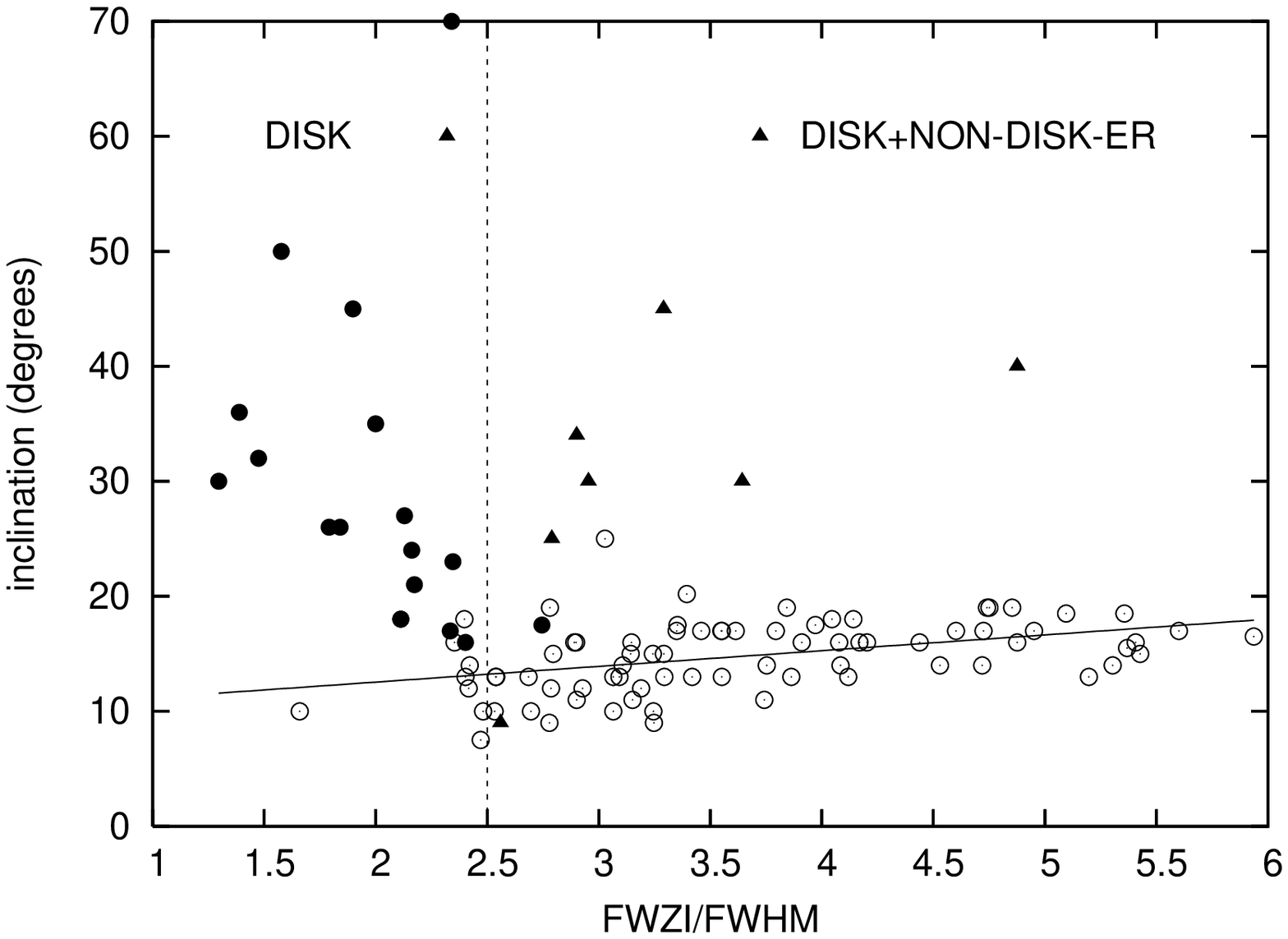}
\includegraphics[width=7 cm]{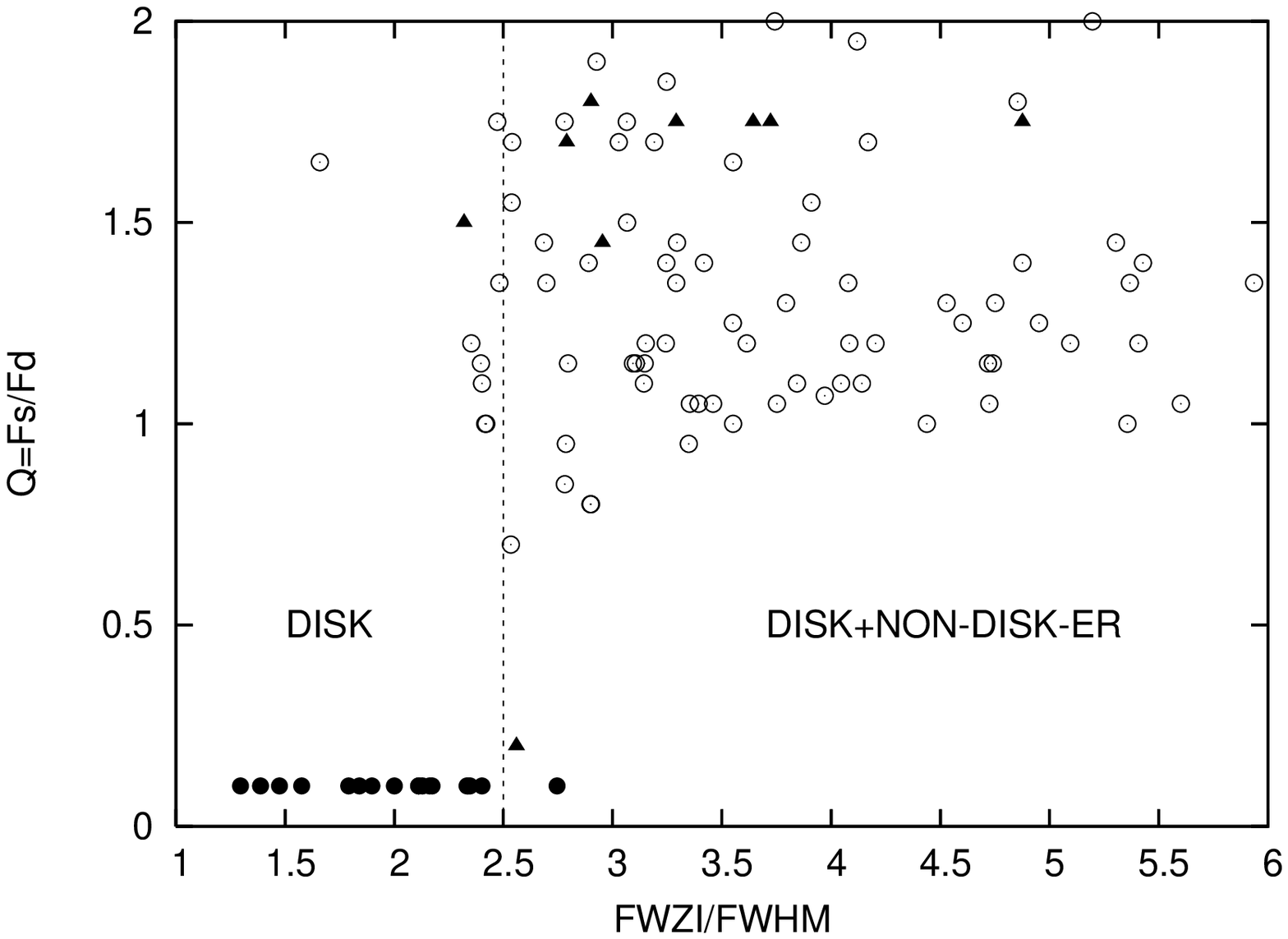}
\caption{Inclination (top) and $Q$ (bottom) as function of the $FWZI/FWHM$. With full 
points the measurements of the disk parameters and widths are taken from Eraclous \& 
Halpern (1994,2003), and open circles are from the sample (measurements of the FWHM and 
FWZI are taken from La Mura et al. 2007). The vertical dashed line shows a rough border between  disk emission lines and ones where the disk may be superposed with the emission of an additional region.} \end{figure}

\section{Discussion}

In the  previous sections we have made a { grid} of  two-component models, aiming
to search for the hidden disk emission. We found that the more significant
parameters in the generation of the emission line profiles are the inclination
and  the flux ratio  between components, $Q$. On the other hand, we have also
explored the influence of the inner radius.  { Comparing  a { grid} of simulated line profiles with the
data of the 90 AGN sample from SDSS, we found that, if the disk emission is
present, the inner radius should not be smaller than 200 R$g$ (it seems to be in the range 1000 R$g>$ R$_{inn}>200$ R$g$)\footnote{ Taking R$_{inn}<200$ R$_g$ or  R$_{inn}>1000$ R$_g$ we could not find a { grid} of models, like shown in Fig. 6, that fit the measured $k_i$ parameters}.}

Fixing the inner and outer radius to an averaged value, obtained from the study
of BELs with two-peaked lines \citep{EH03}, we estimated the values of $i$ and
$Q$ for the 90 AGN sample from SDSS. According to Table 1 {(where we give data for the case without systematic blue shift of the non-disk component)} and the histogram in Figure 9,
the two component model associates significant disk emission to practically all objects. 

{ In Figure 10 we present $Q$ vs $i$, with well known double peaked AGN, assuming that
the contribution of the non-disk region is smaller than 10\%}. {Note here, that even in double-peaked emitters, there is often a residual, so called 'classical broad line' component left over after the disk fit  \citep[as e.g. in 3C390.3, see][]{EH03}.} In this case, there is an indication that the linear regression 
may be present in Q vs. $i$. This may be caused by  the disk brightness, i.e.  with higher inclinations the disk emission  decreases (Eq. 2)
and in this case, the disk emission can be detected if the non-disk emission is negligible.

As an additional test, we have used measurements of the FWHM and FWZI \citep{Gio07} for the 90 AGN sample from SDSS and
for double peaked lines  by  \citet{EH94,EH03} to plot the inclination, $i$, vs. $FWZI/FWHM$. As it can be seen in { Fig. 11}, there is
a clear separation between single and double-peaked lines (except for the points for which we estimate $Q$ and $i$ with high uncertainty, presented as full triangles). This also indicates that a high inclined disk emission can be detected if it is more dominant than {the non-disk component}. {Note here that the lack of AGN population with $Q<1$ (see Fig. 11) is probably caused by selection effects, since we selected a sample where single-peaked profile is dominant.}

The problem of a low-inclined disk ($i<25^\circ$), that we
obtained from this sample of AGN, still remains.
The restriction on inclination
is problematic for disk models, since we may expect random
orientations - at least within the range
allowed by the torus opening angle (given that the accretion disk
and torus are co-axial, see Fig. 12).
Then high-inclined disks may be obscured by the torus, but in this case 
one can expect
the cut-off in the 
inclination around 45$^\circ$-50$^\circ$.
One explanation may be that we have situation as schematically presented in 
Fig. 12,
where the disk is co-axial with  the tours, and there are three cases 
depending on the line-of-sight of an observer;
 1) line-of-sight  is throughout the torus, where the disk and non-disk
regions are obscured and one can detect only narrow lines, 
2) as 
inclination stay smaller,
the non-disk contribution stay important, but the disk emission is still 
full or partly obscured (very close to 
the torus one can expect absorbing material able to absorb disk emission 
and partly non-disk emission); and
3) for a low-inclined disk the absorption coming from the torus is 
negligible, then
disk and non-disk emission can be fully detected.

In the case 2), the fraction of the disk emission may be too weak. 
Consequently, it is very hard to extract possible
disk parameters (see Fig 8, the dots denoted as triangles mainly have 
higher inclinations than 25 degrees).

\section{Conclusions}

Here we investigated the hidden disk emission in single peaked line 
profiles of type 1 AGN in 
order to find any indication that the disk emission is present in non double 
peaked broad lines.

Finally,  we outline the following conclusions:

1) As it was mentioned earlier \citep[]{Pop04,Bon06}, a two component model (disk + non-disk region) can well describe the majority of the observed single peaked line profiles.

2) After comparing simulated and observed line profiles, there is an indication that the disk emission may be present also in single peaked broad line profiles, but it is { mainly} smaller than 50\% of the total line flux.

3) The estimated inclination of the disk indicates a low inclined disk, with inclinations  $i<25^{\circ}$ that might be caused by the torus and/or absorbing material around the torus.

4) In some of  the observed $H{\alpha}$ line profiles there is a blue asymmetry that can be fitted considering a blue-shift  of the non-disk region of  $\sim$800 km/s. This { { may} indicate an outflow in the non-disk region in some AGN from the sample. }

\begin{figure}
\centering
\includegraphics[width=7 cm]{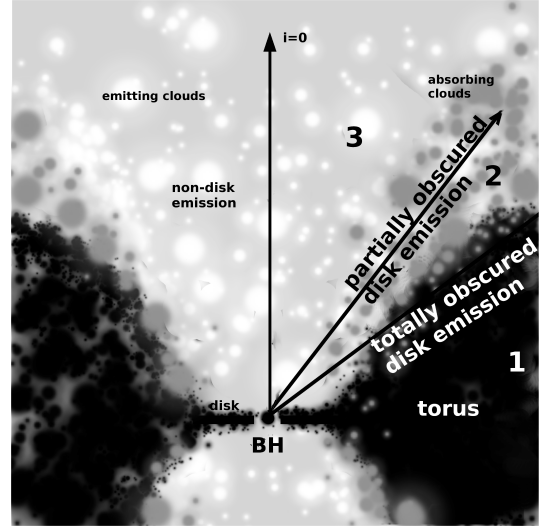}
\caption{Obscuration of the disk emission: 1)torus, 2)absorbing material around the torus and 3) the region without absorption (see text in \S 5).} \end{figure}

\section*{Acknowledgments}

Acknowledgments.
The work was supported by the Ministry of Science
 of Serbia through the project 146002: ``Astrophysical spectroscopy of extragalactic objects``. {We would like to thank to the referee for very useful comments}

\begin{table*}

\centering
 \begin{minipage}{140mm}
  \caption{Measured $k$ and
estimated inclination and ratio of contributions of the non-disk and disk emission.}
  \begin{tabular}{lccccccccc}
\hline

 SDSS name  &  redshit & $k_{10}$ & $k_{20}$ & $k_{30}$ &
$i$ & $\Delta i$ & $ Q $ & $\Delta Q $ \\
\hline
SDSSJ1152-0005	&	0.276	&	3.03	&	2.24	&	1.83	&	19	&	0.05	&	1.05	&	0.05	\\	
SDSSJ1157-0022	&	0.178	&	2.36	&	1.75	&	1.42	&	17	&	0.05	&	1.5	&	0.1	\\	
SDSSJ1307-0036	&	0.188	&	2.90	&	2.11	&	1.61	&	19	&	0.05	&	1.15	&	0.05	\\	
SDSSJ1059-0005	&	0.282	&	2.72	&	1.93	&	1.52	&	18.95	&	0.01	&	1.2	&	0.01	\\	
SDSSJ1342-0053	&	0.129	&	2.76	&	1.91	&	1.48	&	18.95	&	0.01	&	1.2	&	0.01	\\	
SDSSJ1307+0107	&	0.26	&	2.84	&	1.79	&	1.42	&	19	&	0.05	&	1.35	&	0.05	\\	
SDSSJ1341-0053	&	0.17	&	2.52	&	1.79	&	1.42	&	19.35	&	0.1	&	1.5	&	0.1	\\	
SDSSJ1344+0005	&	0.276	&	2.30	&	1.77	&	1.43	&	16.25	&	0.01	&	1.25	&	0.05	\\	
SDSSJ1013-0052	&	0.326	&	2.14	&	1.67	&	1.39	&	17.35	&	3.2	&	1.45	&	0.05	\\	
SDSSJ1010+0043	&	0.237	&	2.30	&	1.80	&	1.50	&	16.8	&	0.45	&	1	&	0.2	\\	
SDSSJ1057-0041	&	0.0871	&	2.03	&	1.60	&	1.33	&	17.25	&	3.0	&	3.3	&	0.3	\\	
SDSSJ0117+0000	&	0.245	&	2.78	&	2.00	&	1.59	&	19	&	0.05	&	1.1	&	0.01	\\	
SDSSJ0112+0003	&	0.0737	&	3.08	&	2.04	&	1.48	&	18.95	&	0.01	&	1.15	&	0.05	\\	
SDSSJ1344-0015	&	0.14	&	2.39	&	1.71	&	1.36	&	18.5	&	0.55	&	2.05	&	0.15	\\	
SDSSJ1343+0004	&	0.114	&	2.48	&	1.84	&	1.48	&	18.25	&	0.1	&	1.15	&	0.15	\\	
SDSSJ1519+0016	&	0.232	&	2.30	&	1.79	&	1.46	&	16.4	&	0.15	&	1.1	&	0.1	\\	
SDSSJ1437+0007	&	0.179	&	2.43	&	1.79	&	1.43	&	17.7	&	0.05	&	1.4	&	0.1	\\	
SDSSJ1619+6202	&	0.31	&	2.18	&	1.59	&	1.31	&	\	&	\	&	\	&	\	\\	
SDSSJ0121-0102	&	0.359	&	2.17	&	1.65	&	1.40	&	15.15	&	0.5	&	1.75	&	0.45	\\	
SDSSJ1719+5937	&	0.174	&	2.55	&	1.80	&	1.40	&	18.95	&	0.01	&	1.5	&	0.01	\\	
SDSSJ1717+5815	&	0.279	&	2.07	&	1.62	&	1.36	&	14.05	&	0.01	&	2.4	&	0.01	\\	
SDSSJ0037+0008	&	0.362	&	2.86	&	2.00	&	1.59	&	18.95	&	0.01	&	1.15	&	0.05	\\	
SDSSJ2351-0109	&	0.252	&	2.07	&	1.67	&	1.38	&	13.1	&	0.05	&	1.3	&	0.1	\\	
SDSSJ2349-0036	&	0.0456	&	2.21	&	1.79	&	1.45	&	15.7	&	0.15	&	0.95	&	0.05	\\	
SDSSJ0013+0052	&	0.239	&	2.31	&	1.71	&	1.43	&	16.45	&	0.1	&	1.45	&	0.25	\\	
SDSSJ1720+5540	&	0.0543	&	2.14	&	1.59	&	1.33	&	\	&	\	&	\	&	\	\\	
SDSSJ0256+0113	&	0.0804	&	1.75	&	1.52	&	1.31	&	13.2	&	3.95	&	0.3	&	0.01	\\	
SDSSJ0135-0044	&	0.334	&	1.94	&	1.56	&	1.33	&	\	&	\	&	\	&	\	\\	
SDSSJ0140-0050	&	0.146	&	2.66	&	1.88	&	1.50	&	19	&	0.05	&	1.25	&	0.05	\\	
SDSSJ0310-0049	&	0.217	&	2.13	&	1.70	&	1.39	&	14.05	&	0.01	&	1.3	&	0.01	\\	
SDSSJ0304+0028	&	0.368	&	1.95	&	1.67	&	1.36	&	12.95	&	2.3	&	0.9	&	0.4	\\	
SDSSJ0159+0105	&	0.198	&	2.54	&	1.81	&	1.46	&	18.85	&	0.1	&	1.35	&	0.15	\\	
SDSSJ0233-0107	&	0.177	&	2.30	&	1.78	&	1.46	&	16.4	&	0.15	&	1.1	&	0.1	\\	
SDSSJ0250+0025	&	0.0445	&	2.71	&	1.76	&	1.41	&	19.05	&	0.01	&	1.4	&	0.1	\\	
SDSSJ0409-0429	&	0.0802	&	2.30	&	1.76	&	1.43	&	16.3	&	0.05	&	1.25	&	0.15	\\	
SDSSJ0937+0135	&	0.107	&	2.35	&	1.72	&	1.41	&	16.85	&	0.1	&	1.55	&	0.15	\\	
SDSSJ0323+0035	&	0.185	&	2.69	&	1.88	&	1.50	&	19	&	0.05	&	1.25	&	0.05	\\	
SDSSJ0107+1408	&	0.215	&	2.95	&	1.95	&	1.50	&	18.95	&	0.01	&	1.2	&	0.01	\\	
SDSSJ0142+0005	&	0.0768	&	2.53	&	1.80	&	1.40	&	19.1	&	0.15	&	1.55	&	0.05	\\	
SDSSJ0306+0003	&	0.0941	&	2.95	&	1.94	&	1.44	&	18.95	&	0.01	&	1.2	&	0.01	\\	
SDSSJ0322+0055	&	0.0893	&	2.95	&	2.00	&	1.44	&	19	&	0.05	&	1.2	&	0.01	\\	
SDSSJ0150+1323	&	0.0371	&	2.63	&	2.04	&	1.67	&	19.05	&	0.01	&	0.95	&	0.05	\\	
SDSSJ0855+5252	&	0.0691	&	2.75	&	1.87	&	1.44	&	19.05	&	0.01	&	1.3	&	0.01	\\	
SDSSJ0904+5536	&	0.0388	&	2.41	&	1.81	&	1.44	&	17.45	&	0.01	&	1.25	&	0.05	\\	
SDSSJ1355+6440	&	0.0505	&	2.80	&	1.95	&	1.50	&	18.95	&	0.01	&	1.2	&	0.01	\\	
SDSSJ0351-0526	&	0.0751	&	2.29	&	1.68	&	1.38	&	18.4	&	1.75	&	1.95	&	0.05	\\	
SDSSJ1505+0342	&	0.0583	&	2.67	&	1.81	&	1.48	&	18.95	&	0.01	&	1.3	&	0.1	\\	
SDSSJ1203+0229	&	0.0931	&	2.30	&	1.73	&	1.38	&	16.7	&	0.35	&	1.8	&	0.2	\\	
SDSSJ1246+0222	&	0.0775	&	2.11	&	1.63	&	1.37	&	14.1	&	0.25	&	2	&	0.2	\\	
SDSSJ0839+4847	&	0.0236	&	2.13	&	1.68	&	1.40	&	14.15	&	0.1	&	1.3	&	0.1	\\	
SDSSJ0925+5335	&	0.0867	&	2.78	&	1.89	&	1.44	&	19	&	0.05	&	1.25	&	0.05	\\	
SDSSJ1331+0131	&	0.0482	&	2.94	&	1.78	&	1.39	&	19	&	0.05	&	1.35	&	0.05	\\	
SDSSJ1042+0414	&	0.0805	&	6.82	&	1.58	&	1.36	&	19.05	&	0.01	&	1.3	&	0.01	\\	
SDSSJ1349+0204	&	0.0328	&	2.55	&	1.92	&	1.49	&	19.05	&	0.1	&	1.1	&	0.1	\\	
SDSSJ1223+0240	&	0.0722	&	2.17	&	1.69	&	1.41	&	14.75	&	0.1	&	1.3	&	0.2	\\	
SDSSJ0755+3911	&	0.0335	&	2.41	&	1.77	&	1.45	&	17.5	&	0.05	&	1.3	&	0.2	\\	
SDSSJ1141+0241	&	0.0459	&	2.78	&	1.96	&	1.52	&	19	&	0.05	&	1.2	&	0.01	\\	
SDSSJ1122+0117	&	0.0394	&	2.70	&	1.85	&	1.45	&	19	&	0.05	&	1.25	&	0.05	\\	
SDSSJ1243+0252	&	0.0767	&	2.50	&	1.81	&	1.50	&	18.85	&	0.5	&	1.2	&	0.2	\\	
SDSSJ0832+4614	&	0.0605	&	3.00	&	1.90	&	1.41	&	19	&	0.05	&	1.25	&	0.05	\\	
SDSSJ0840+0333	&	0.0525	&	2.04	&	1.53	&	1.33	&	\	&	\	&	\	&	\	\\	
SDSSJ1510+0058	&	0.0359	&	2.35	&	1.79	&	1.50	&	17.05	&	0.3	&	1.1	&	0.2	\\	

\hline
\end{tabular}
\end{minipage}
\end{table*}

\addtocounter{table}{-1}
\begin{table*}

\centering
 \begin{minipage}{140mm}
  \caption{continued}
  \begin{tabular}{lccccccccc}
\hline

SDSSJ0110-1008	&	0.078	&	2.39	&	1.82	&	1.49	&	17.45	&	0.1	&	1.15	&	0.15	\\	
SDSSJ0142-1008	&	0.0303	&	2.33	&	1.63	&	1.32	&	19	&	0.05	&	2.4	&	0.01	\\	
SDSSJ1519+5208	&	0.0693	&	2.53	&	1.74	&	1.42	&	18.85	&	0.1	&	1.6	&	0.1	\\	
SDSSJ0013-0951	&	0.0738	&	2.53	&	1.80	&	1.53	&	19.1	&	0.15	&	1.2	&	0.3	\\	
SDSSJ1535+5754	&	0.0615	&	2.12	&	1.64	&	1.36	&	15.5	&	1.15	&	2.5	&	0.5	\\	
SDSSJ1654+3925	&	0.0419	&	2.23	&	1.67	&	1.37	&	17.05	&	0.9	&	2.3	&	0.3	\\	
SDSSJ0042-1049	&	0.0581	&	2.52	&	1.78	&	1.43	&	19.15	&	0.3	&	1.45	&	0.15	\\	
SDSSJ2058-0650	&	0.0904	&	2.50	&	1.81	&	1.38	&	19.25	&	0.2	&	1.65	&	0.15	\\	
SDSSJ1300+6139	&	0.0522	&	2.10	&	1.73	&	1.42	&	14.55	&	0.2	&	0.85	&	0.05	\\	
SDSSJ0752+2617	&	0.0948	&	2.74	&	1.87	&	1.48	&	19	&	0.05	&	1.25	&	0.05	\\	
SDSSJ1157+0412	&	0.0822	&	2.84	&	1.89	&	1.42	&	19	&	0.05	&	1.25	&	0.05	\\	
SDSSJ1139+5911	&	0.0854	&	2.30	&	1.70	&	1.39	&	18.55	&	1.7	&	1.75	&	0.15	\\	
SDSSJ1345-0259	&	0.0279	&	2.00	&	1.53	&	1.30	&	20.45	&	0.01	&	3.7	&	0.1	\\	
SDSSJ1118+5803	&	0.0613	&	2.32	&	1.73	&	1.46	&	16.65	&	0.1	&	1.3	&	0.3	\\	
SDSSJ1105+0745	&	0.0734	&	2.45	&	1.68	&	1.43	&	18.45	&	0.6	&	1.65	&	0.35	\\	
SDSSJ1623+4804	&	0.0449	&	2.33	&	1.71	&	1.37	&	20	&	0.15	&	2	&	0.2	\\	
SDSSJ0830+3405	&	0.0696	&	2.00	&	1.60	&	1.35	&	12.1	&	0.15	&	2.05	&	0.15	\\	
SDSSJ1619+4058	&	0.0335	&	2.12	&	1.62	&	1.34	&	17.75	&	1.6	&	3.1	&	0.3	\\	
SDSSJ0857+0528	&	0.0379	&	2.39	&	1.77	&	1.42	&	17.2	&	0.05	&	1.4	&	0.1	\\	
SDSSJ1613+3717	&	0.0586	&	2.49	&	1.77	&	1.47	&	19.05	&	0.6	&	1.35	&	0.25	\\	
SDSSJ1025+5140	&	0.0623	&	3.00	&	2.11	&	1.54	&	19	&	0.05	&	1.15	&	0.05	\\	
SDSSJ1016+4210	&	0.0553	&	2.63	&	1.83	&	1.42	&	19.05	&	0.01	&	1.35	&	0.05	\\	
SDSSJ1128+1023	&	0.0504	&	2.88	&	2.00	&	1.53	&	18.95	&	0.01	&	1.15	&	0.05	\\	
SDSSJ1300+5641	&	0.0718	&	2.75	&	2.00	&	1.58	&	19	&	0.05	&	1.1	&	0.01	\\	
SDSSJ1538+4440	&	0.0406	&	2.29	&	1.79	&	1.45	&	18.2	&	2.05	&	1.1	&	0.1	\\	
SDSSJ1342+5642	&	0.0728	&	2.46	&	1.88	&	1.54	&	18.45	&	0.4	&	0.95	&	0.15	\\	
SDSSJ1344+4416	&	0.0547	&	2.68	&	1.89	&	1.42	&	19	&	0.05	&	1.25	&	0.05	\\	
SDSSJ1554+3238	&	0.0483	&	2.03	&	1.48	&	1.26	&	20.4	&	0.05	&	3.4	&	0.01	\\	
\hline
\end{tabular}
\end{minipage}
\end{table*}

\appendix

\section[]{The cases of the different Gaussian widths}

{

As we mentioned  in \S 2.2, we fixed the widths of Gaussian (that represents the non-disk component) to $\sigma=$1000 km/s.
Recall the results of \citet{Pop04} and \citet{Bon06}, where the $\sigma$ of the Doppler broadening from the non disk component (the Gaussian representing ILR) were estimated to be from 300 to 1700 km/s. This is in a good agreement with the results of \citet{Hu08}. They used a two component model assuming that a broad line can be represented by two Gaussians (one very broad, and another with intermediate velocities, corresponding to ILR and VBLR respectively) that FWHM's from VBLR and ILR components are in correlation, as FWHM(ILR) $\sim$  0.4 FWHM(VLBR), and FWHM (ILR) ranged from 1000 to 4000 km/s ($\sigma$  $\sim$ 400 to $\sim$ 1700 km/s) \citep{Hu08}.

But, to have an impression how $\sigma$ of the non-disk component can affect composite profiles, we performed simulations using  different $\sigma$. We started from $\sigma$  $\sim$ 400 to $\sim$ 1700 km/s which corresponds FWHM $\sim$ 1000 to $\sim$  4000 km/s, with a step of $\Delta$ FWHM = 1000 km/s.  The disk parameters are taken as R$_{inn}$=600 R$g$, R$_{out}$=4000 R$g$, emissivity $p=3$, and we considered different inclinations and $Q$. In Figs. \ref{appenfig1}-\ref{appenfig3} we presented calculated parameters $k_{20},\ k_{30}$ vs. $k_{10}$ for Q=0.3, 1 and 2, taking $\sigma =$ 500 km/s, 1000 km/s and 1700 km/s and different inclinations. Also, the elliptical surface shown in Figs. \ref{appenfig1}-\ref{appenfig3} presents the surface where the measured data from the sample are located. As one can see, different values of $\sigma$ can affect obtained results for Q and inclination, but it is interesting that measured data indicates  $\sigma$  $\sim$ 1000 km/s well fit the surface of measured values in all three considered cases.  Also, for the very broad non-disk component ($\sigma>1700$ km/s) it is hard to use this model to estimate the disk parameters (especially for small inclinations).


}

\begin{figure*}
\centering
\includegraphics[width=7.5 cm, angle=270]{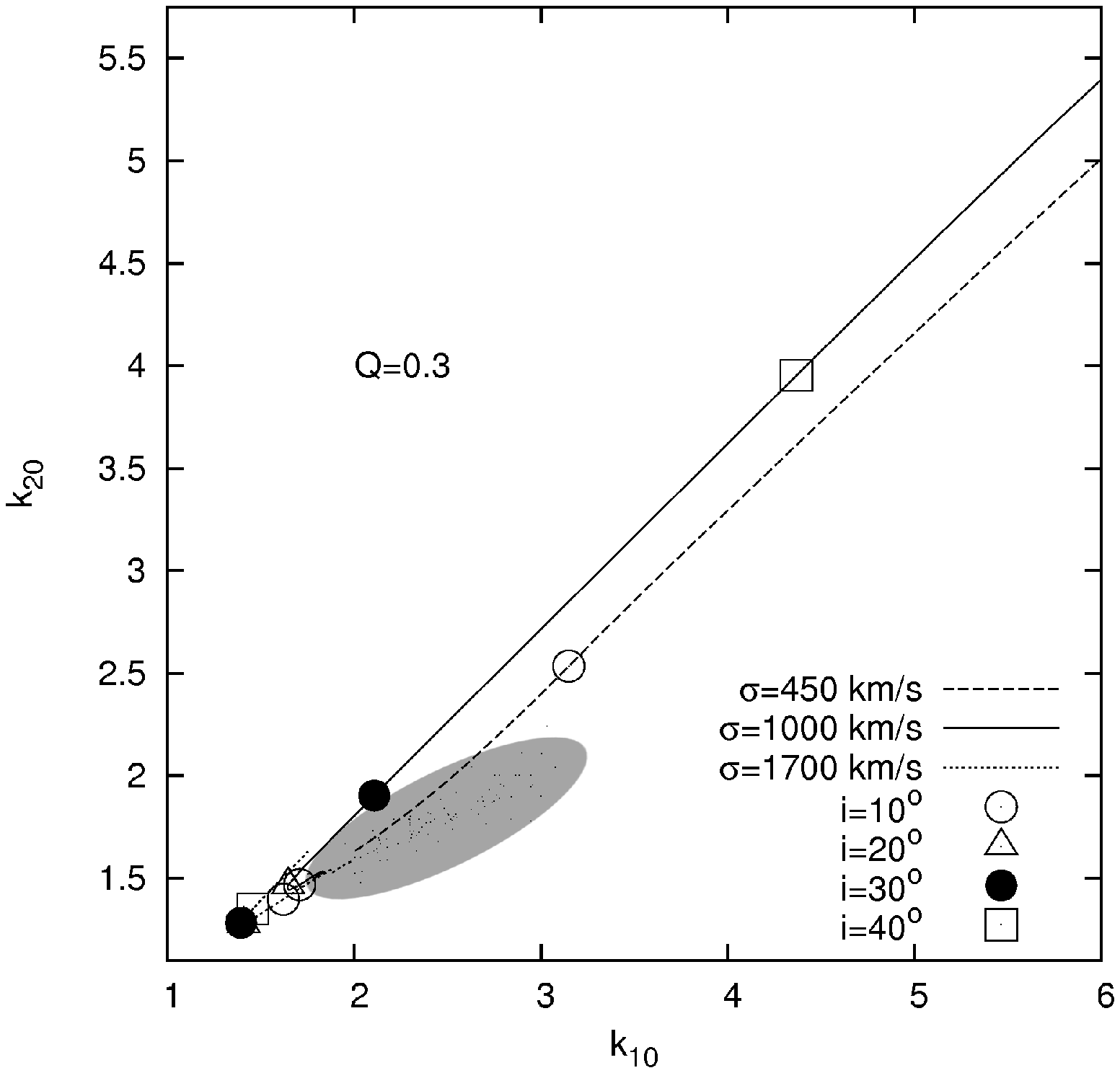}
\includegraphics[width=7.5 cm, angle=270]{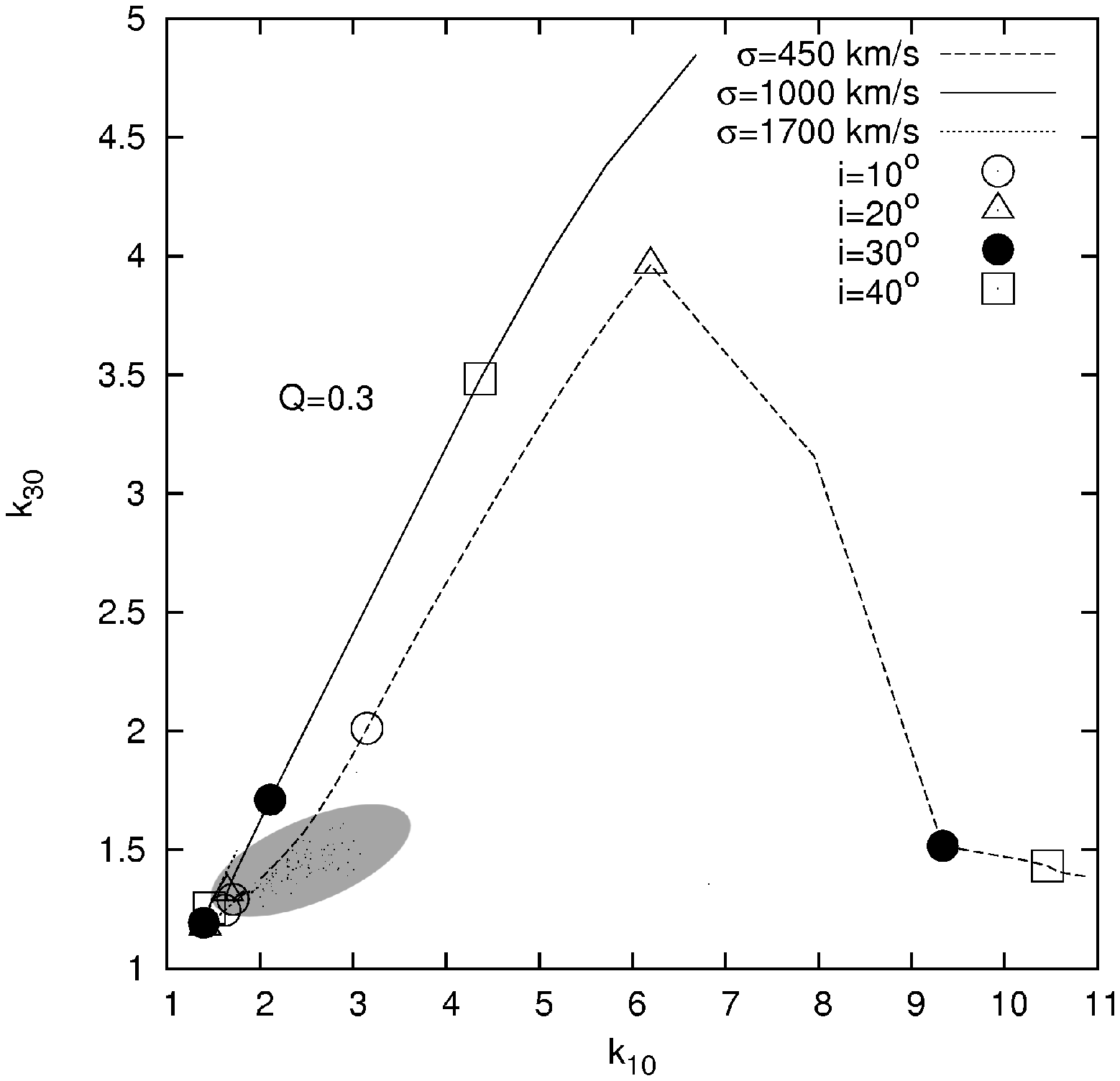}
\caption{{
Simulated $k_{20}$ vs. $k_{10}$ (left) and  $k_{30}$ vs. $k_{10}$ (right) for different $\sigma$ for the non-disk component. The elliptical 
surface denotes the surface of measured values from the sample. The disk parameters are taken as: Rinn=600 Rg, Rout=4000 Rg, $\sigma_{d}$= 1000 km/s}}
\label{appenfig1}
\end{figure*}

\begin{figure*}
\centering
\includegraphics[width=7.5 cm, angle=270]{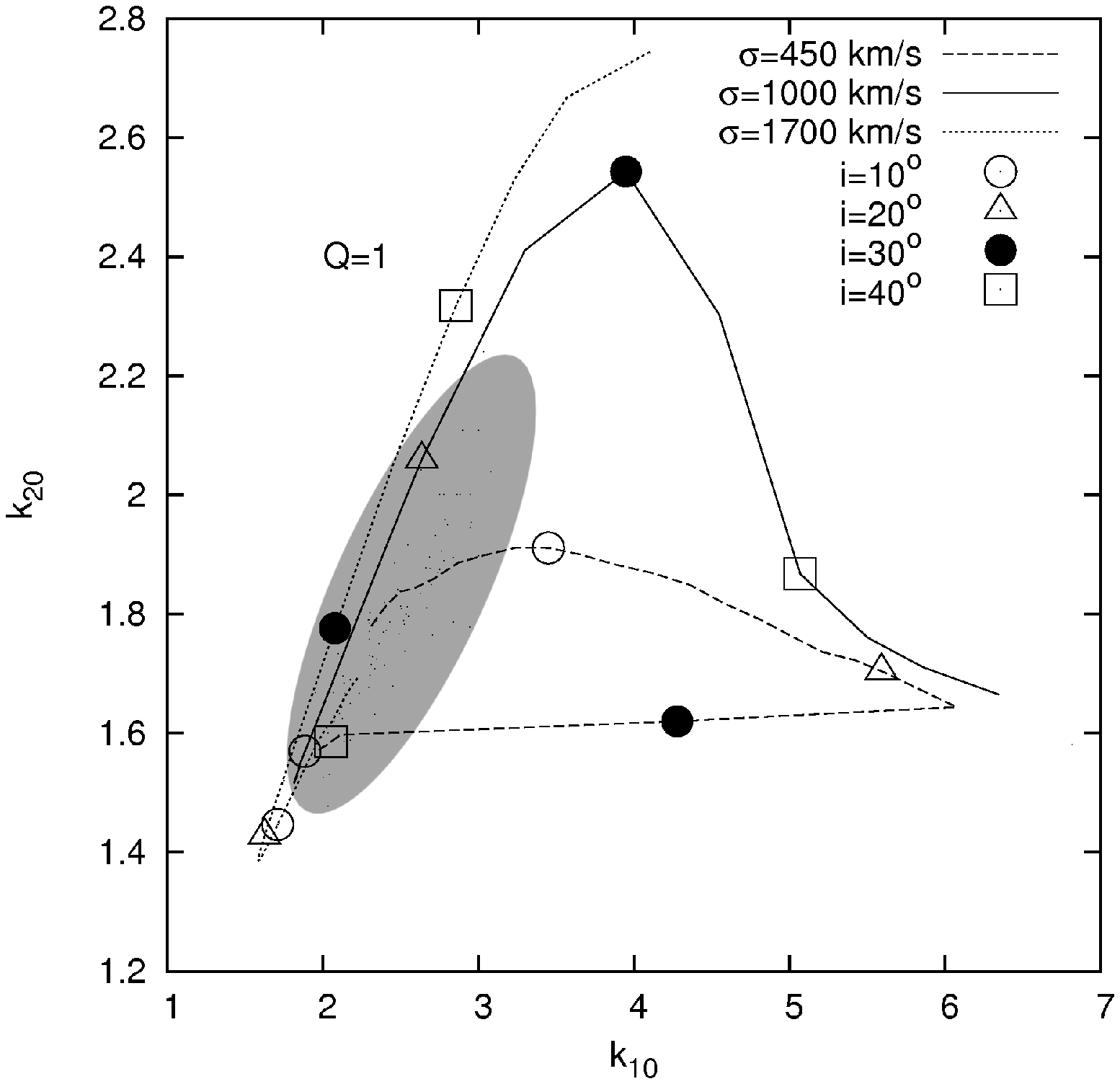}
\includegraphics[width=7.5 cm, angle=270]{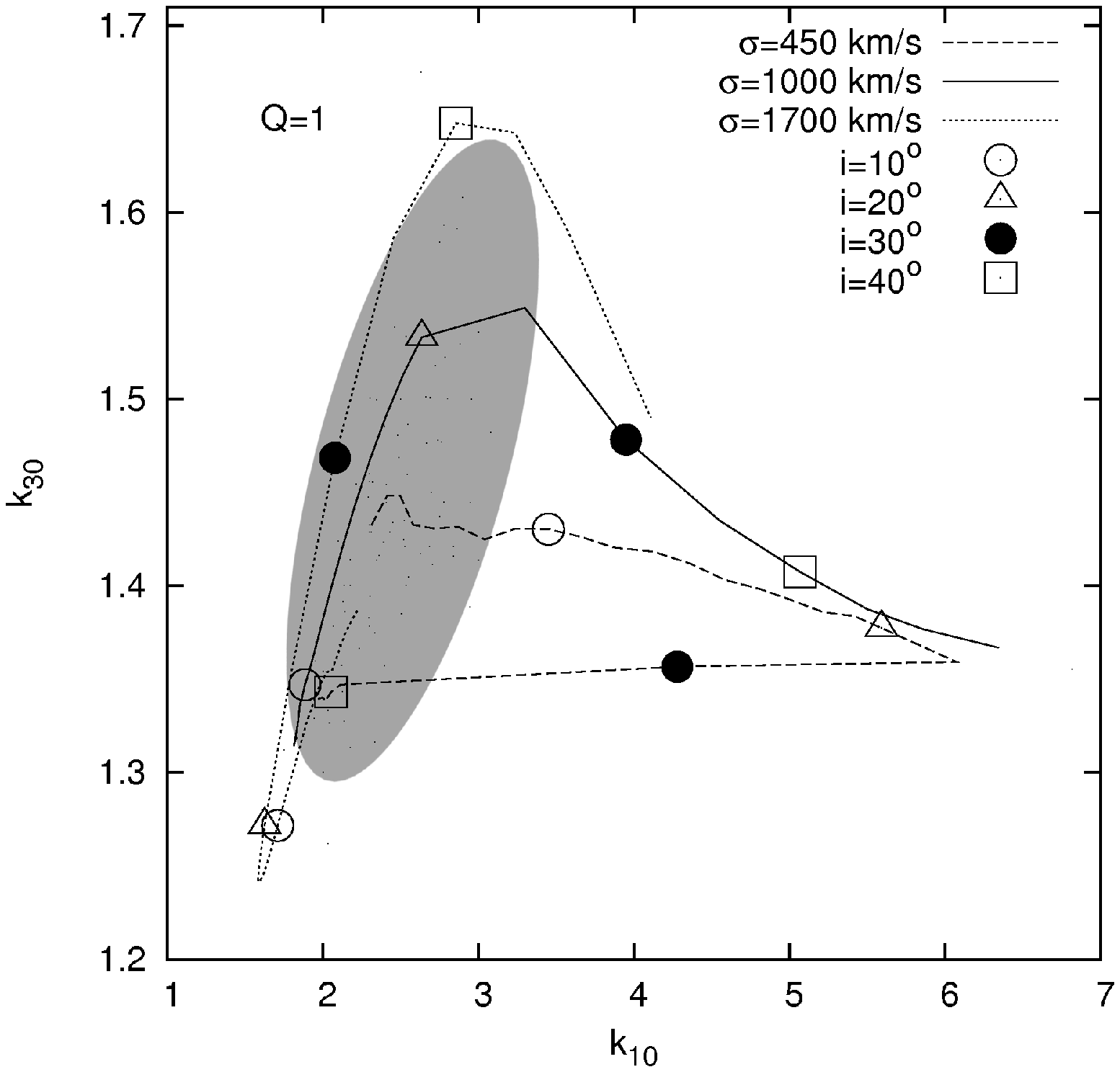}
\caption{{
The same as in Fig 13, but for Q=1.}}
\label{appenfig2}
\end{figure*}

\begin{figure*}
\centering
\includegraphics[width=7.5 cm, angle=270]{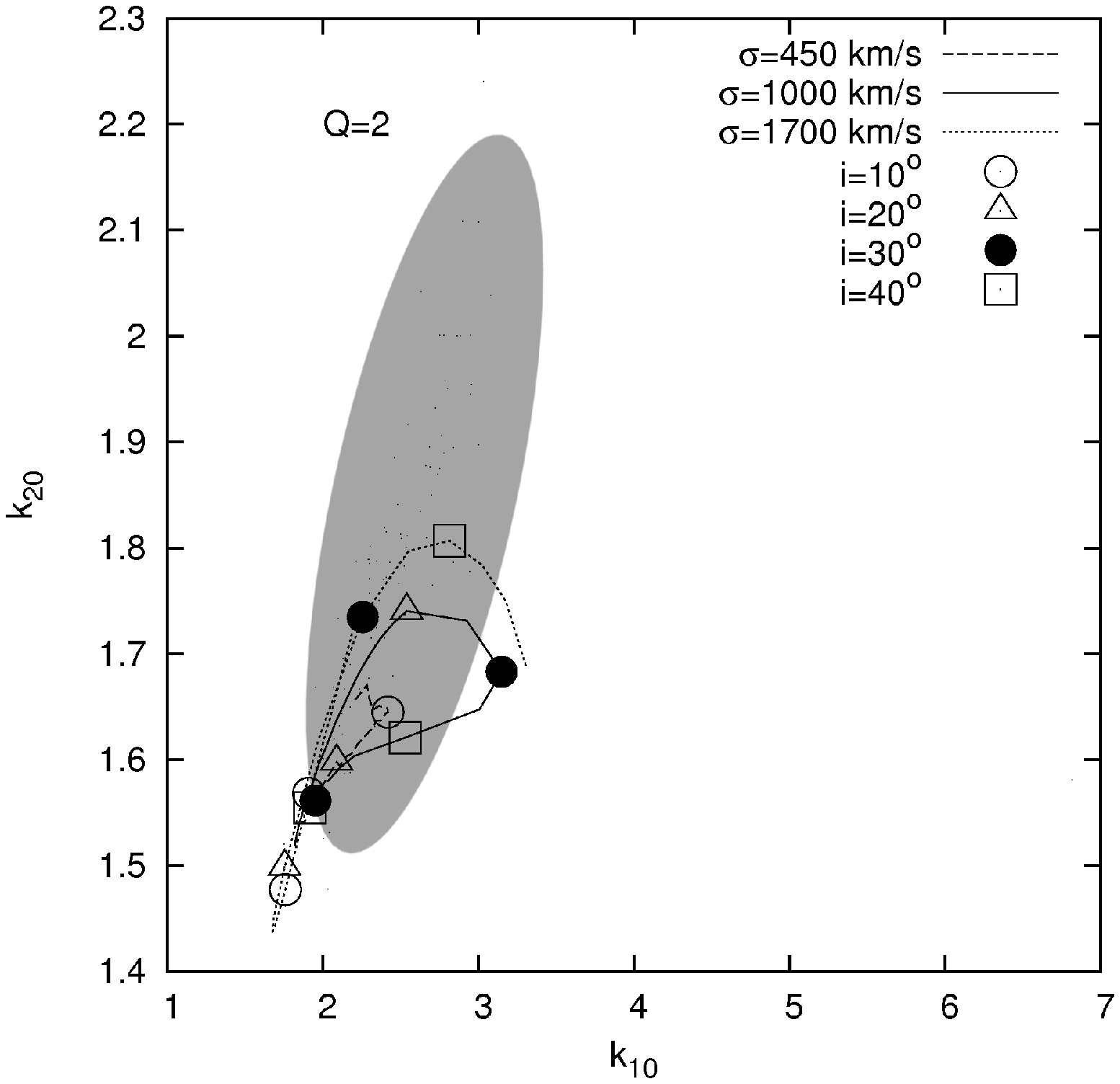}
\includegraphics[width=7.5 cm, angle=270]{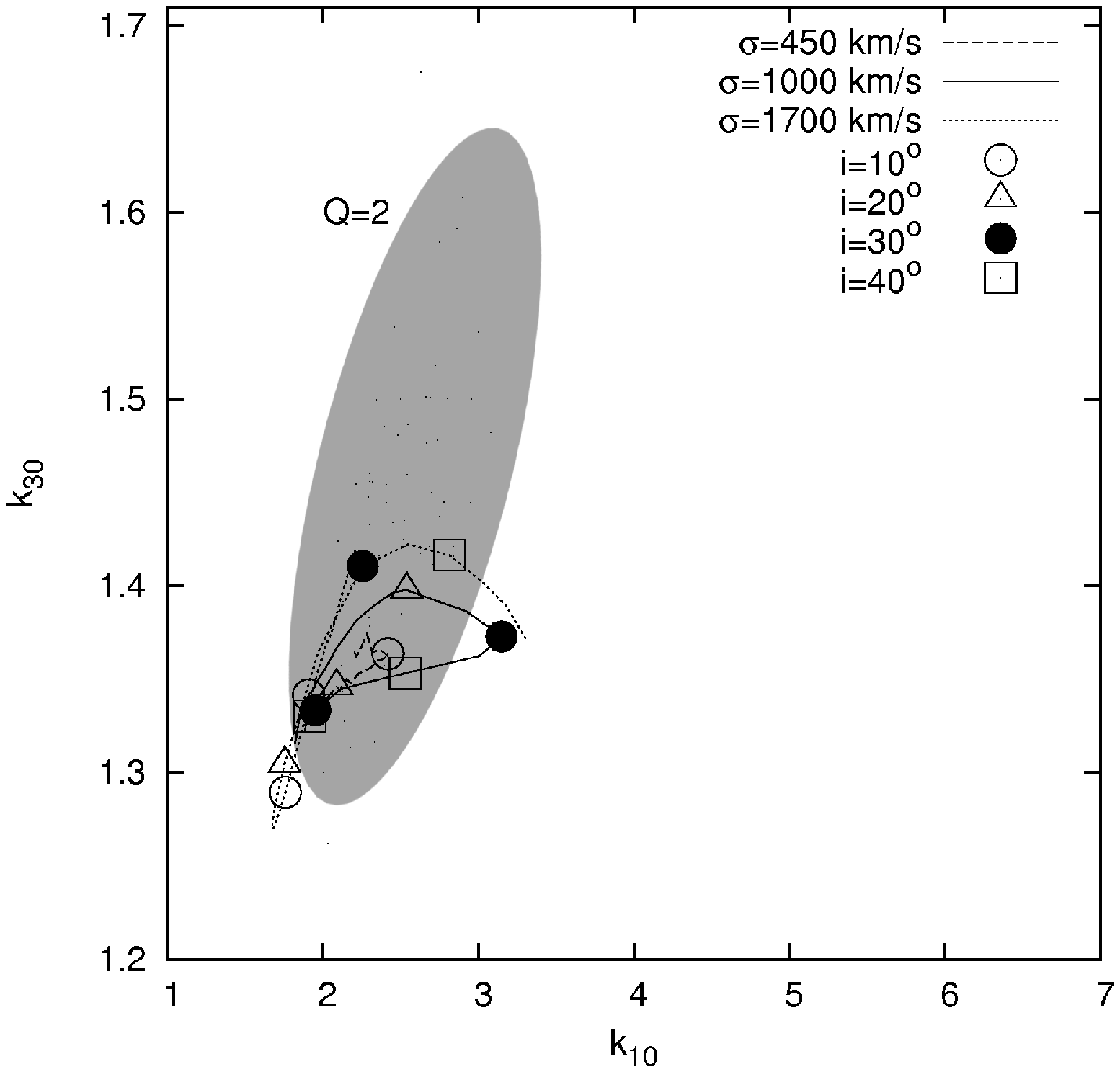}
\caption{{The same as in Fig 13, but for Q=2.}}
\label{appenfig3}
\end{figure*}

\section[]{The two-component model vs. two-Gaussian fitting}
{
As we mentoned above, there are several possibilities for the BLR geometry, i.e. for 
geometry of the VBLR and ILR. In principle, there are many analyzes based on two broad Gaussian
in order to explain physics and geometry of the BLR \citep[see e.g.][]{mar09}.

Here we briefly discuss the fitting using the two-component model \citep[as it was described in][]{Pop04} and two Gaussians. As an example here we show the best fit with the two-component model  (Fig. \ref{appenfigb2}a, left) and with the two-Gaussian one (Fig. \ref{appenfigb2}b). As one can see in Fig. \ref{appenfigb2}, both models can well fit the complex line profiles. The obtained kinematical parameters are: 1) two-Gaussian FWHM$_{VBLR}=$3760 km/s (redshifted $\sim$560 km/s), FWHM$_{ILR}=$1390 km/s (in the center) and Q=F$_{ILR}$/F$_{VBLR}\approx$0.6; 2) two-component model with the disk emission  FWHM$_{Disk}=$1620 km/s (redshifted $\sim$0 km/s), FWHM$_{ILR}=$1360 km/s
and Q=F$_{ILR}$/F$_{Disk}\approx$0.8. 
The obtained kinematical parameters in both cases indicate that VBLR has higher velocities, but in the case of the assumed disk emission the random velocity is comparable with one present in the ILR.

Comparing the VBLR Gaussian and disk component we found that mainly the VBLR Gaussian should be shifted to the red in order to fit complex BEL profiles \citep[similar as in the case of so called Pop B objects, see][]{sul09}, while the disk component is more consistent with the ILR component.

\begin{figure*}
\centering
\includegraphics[width=7.5 cm, angle=0]{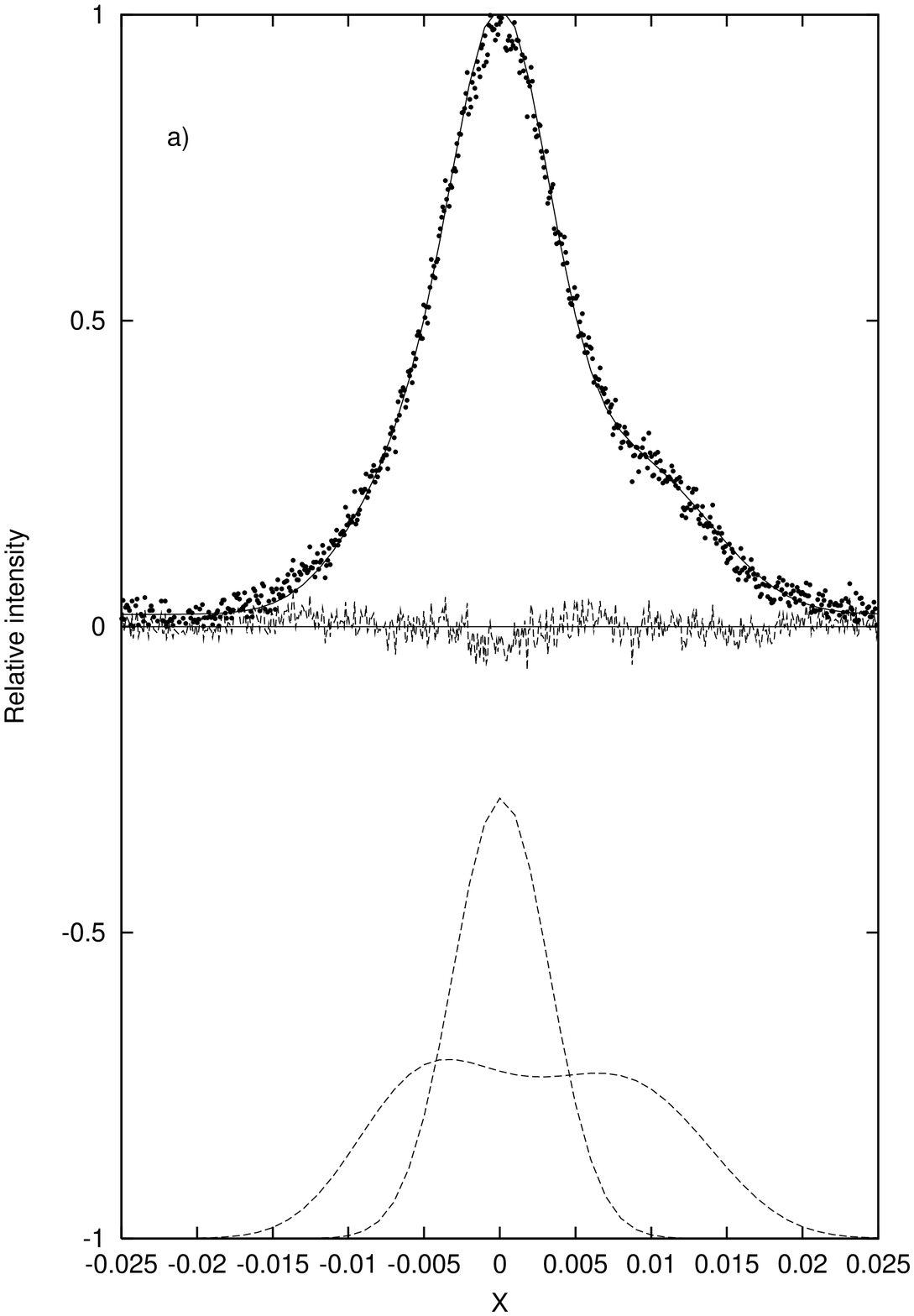}
\includegraphics[width=7.5 cm, angle=0]{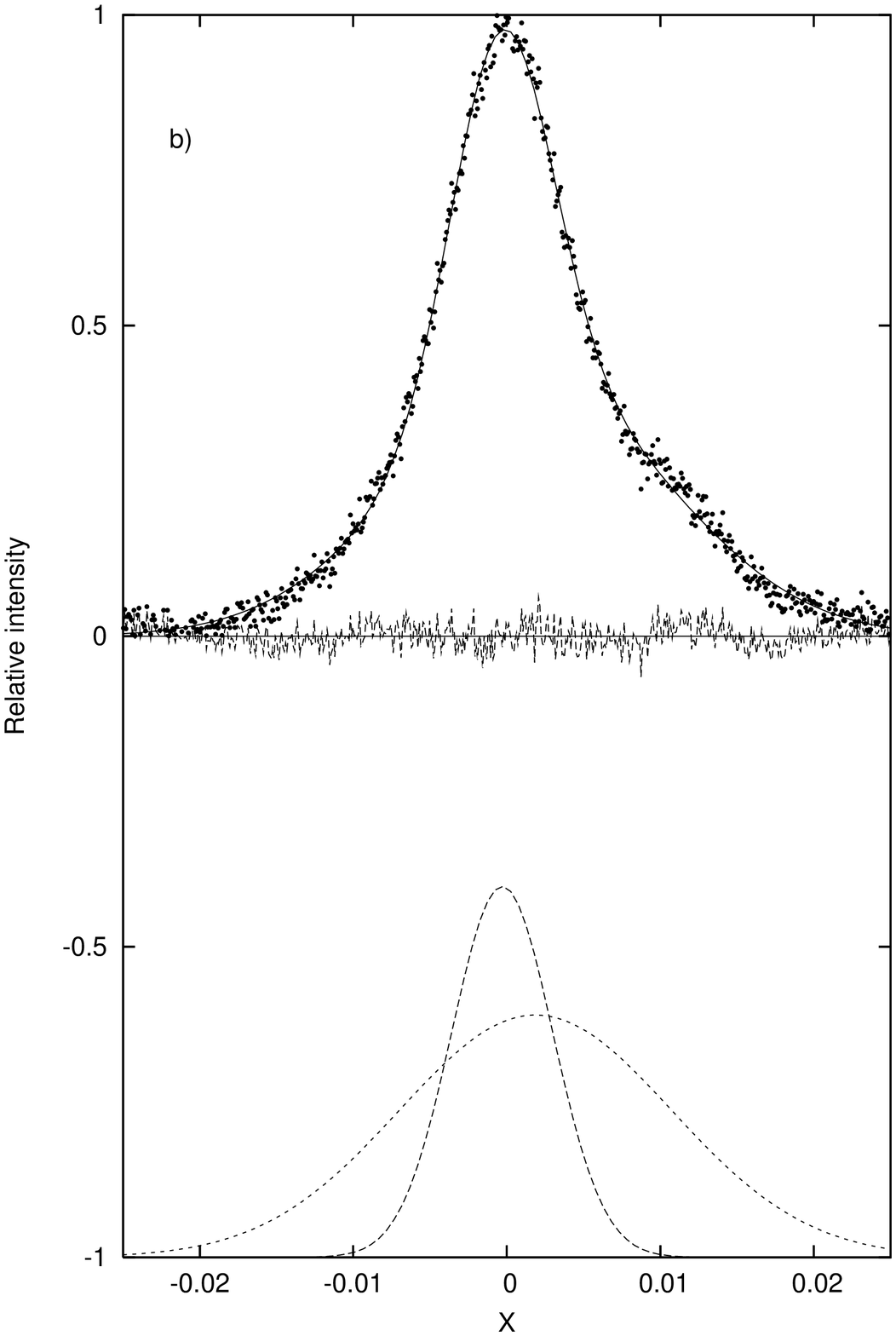}
\caption{The best fit with two broad Gaussian (left) and with two-component model (right) of 3C273 \citep[see][]{Pop04} where one component is assumed to be emitted from the disk (a double peaked component below).}
\label{appenfigb2}
\end{figure*}

Additionally, we did several tests taking into account the two-Gaussian model, fixing one that represents the ILR emission at FWHM=1000 km/s, and changing the width of the one representing the VBLR from 1000 km/s to 10000 km/s taking different Q=F$_{ILR}$/F$_{VBLR}$=0.3, 0.5,  1,  and 2 (see Fig. \ref{appenfig1b}). We also measured $k_i$ for such modeled profiles and compared them with measured from observed profiles (Fig. \ref{appenfig1b}). In Fig. \ref{appenfig1b} the lines represent modeled values, and crosses measured ones.  As it can be seen 
such model may describe majority of the observed profiles, but there is a big difference between widths for the VBLR component estimated using $k_{20,30}$ vs. $k_{10}$. Also, in this case we obtain that larger fraction of observed AGN has Q=F$_{ILR}$/F$_{VBLR}<$1, i.e. that the VBLR component is dominant in line profiles.
Comparing these two models (see Figs. 6 and \ref{appenfig1b}), we found that the two-component model with the disk emission gives more consistent results (Q and $i$, see Table 1)

\begin{figure*}
\centering
\includegraphics[width=7.5 cm, angle=270]{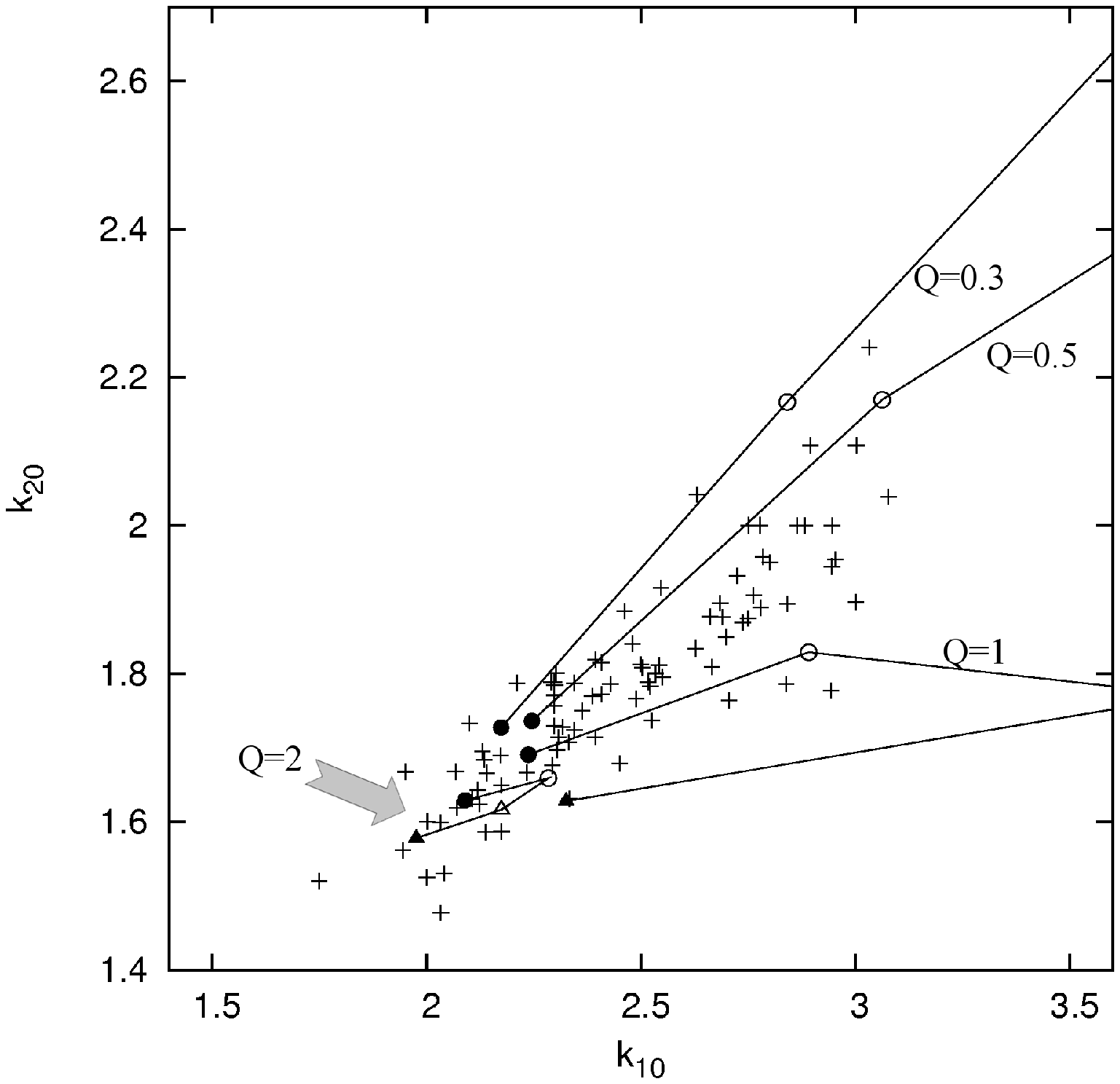}
\includegraphics[width=7.5 cm, angle=270]{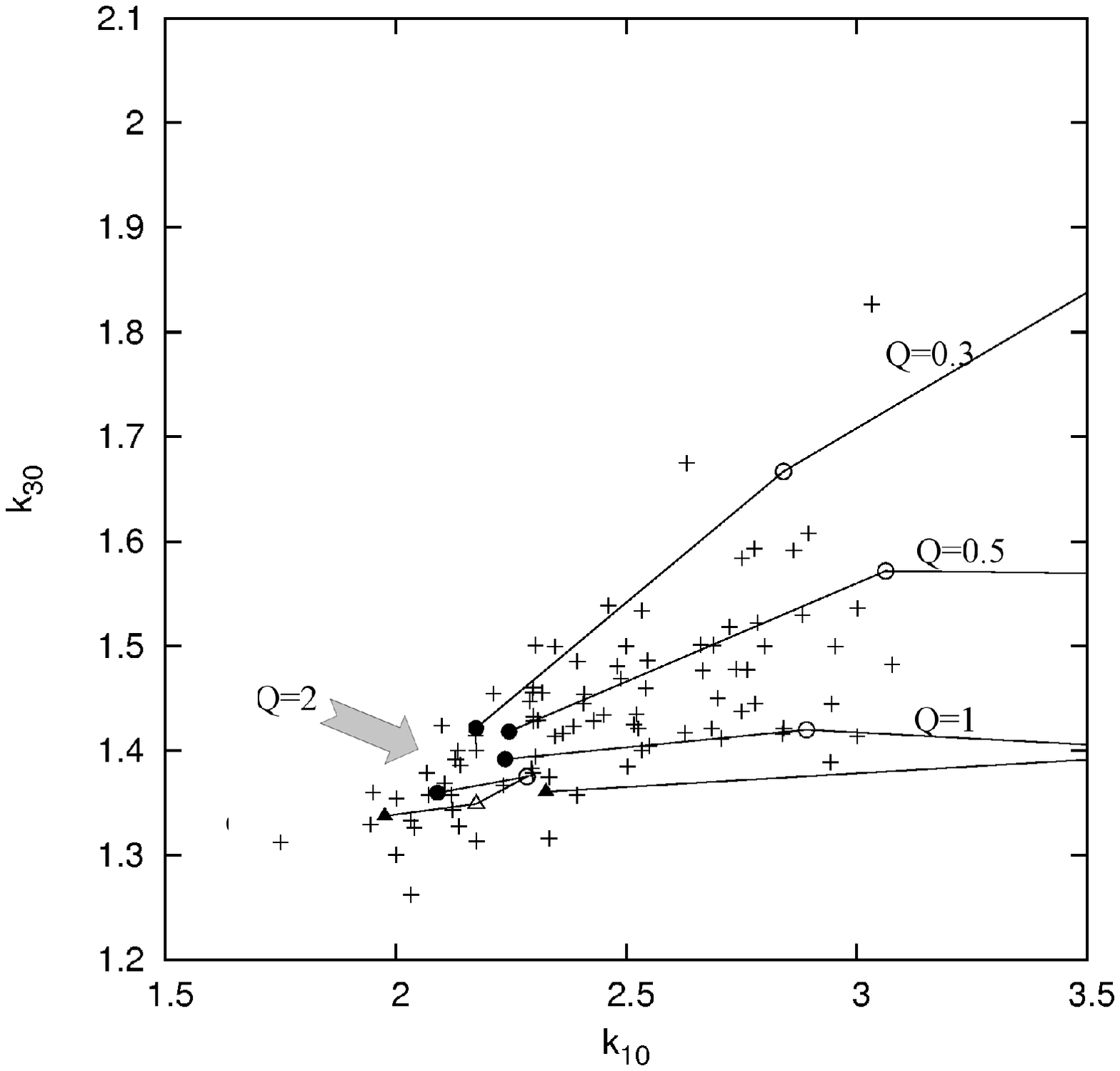}
\caption{
The measured width ratios  (crosses) and simulated 
values (lines) from the two-Gaussian model for the different  
contribution of the disk 
emission to the total line flux ($Q=F_{ILR}/F_{VBLR}$=0.3, 0.5,  1,  and 2). The width of the ILR component is fixed as 1000 km/s, and with the VBLR component is taken as 2000 km/s (full circles),  3000 km/s (open circles), 5000 km/s (open triangles) and 10000 km/s (full triangles).  
}
\label{appenfig1b}
\end{figure*}

In principle for some kind of investigation it is useful (and quit simpler) to use the two-Gaussian fit, but here we aim to investigate possible presence of the disk emission in single-peaked lines, and physically the two-component model with a disk emission seems to better explain the nature of AGN. }



\bsp

\label{lastpage}


\begin{thebibliography}{99}
\bibitem[\protect\citeauthoryear{Baird}{1981}]{b1} Baird S.R., 1981, ApJ, 245, 208.

\bibitem[\protect\citeauthoryear{Bon et al.} {2006}]{Bon06} Bon, E., Popovi\'c, L. \v C.,  Ili\'c, D., Mediavlilla, E.G., 2006,
NewAstRev, 50, 716.

\bibitem[\protect\citeauthoryear{Bon} {2008}]{Bon08} Bon, E., 2008. SerAJ, 177.

\bibitem[\protect\citeauthoryear{Cao \& Wang} {2006}]{caowang05} Cao, X and Wang, T.-G. 2006, ApJ, 652, 112

\bibitem[\protect\citeauthoryear{Chen \& Halpern}{1989}] {Chen89b} Chen, K. \& Halpern, J.P. 1989, ApJ, 344, 115.

\bibitem[\protect\citeauthoryear{Chen et al.} {1989}]{Chen89a} Chen, K., Halpern, J.P. \& Filippenko, A.V. 1989, ApJ, 339, 742.

\bibitem[\protect\citeauthoryear{Collin et al.}{2006}] {Col06} Collin, S., Kawaguchi, T., Peterson, B. M., Vestergaard, M. 2006, A\&AS, 456, 75


\bibitem[\protect\citeauthoryear{Corbin \& Boroson} {1996}]{CB96} Corbin, M. R. \& Boroson, T. A. 1996, ApJS, 107, 69.

\bibitem[\protect\citeauthoryear{Dumont \& Collin-Souffrin}{1990}]{Dum90} Dumont, A.M. \& Collin-Souffrin, S. 1990, A{\&}AS 83, 71.

\bibitem[\protect\citeauthoryear{Eracleous \& Halpern} {1994}]{EH94} Eracleous, M. \& Halpern, J.P. 1994, ApJS, 90, 1.

\bibitem[\protect\citeauthoryear{Eracleous \& Halpern} {2003}]{EH03} Eracleous, M. \& Halpern, J.P. 2003, ApJ, 599, 886.

\bibitem[\protect\citeauthoryear{Ho et al.} {2000}]{Ho00} Ho, L.C., Rudnick, G., Rix, H.-W. et al. 2000, ApJ 541, 120.

\bibitem[\protect\citeauthoryear{Hu et al.} {2008}]{Hu08} Hu, C. , Wang, J. M., Chen, Y. M., Bian W. H., Xue S. J., 2008, ApJ, 683L, 115H

\bibitem[\protect\citeauthoryear{Ili\'c et al.} {2006}]{Ilic06} Ili\'c, D., Popov\'c, L.\v C., Bon, E. Mediavilla, E. G,, Chavushyan, V. H. 2006, MNRAS, 371, 1610

\bibitem[\protect\citeauthoryear{Jovanovi\'c \& Popovi\'c} {2008}]{jov08} Jovanovi\'c, P. \& Popovi\'c L.\v C., 2008, Fortschr. Phys. 56, No. 4 - 5, 456.

\bibitem[\protect\citeauthoryear{Kollatschny \& Bischoff} {2002}]{Koll02} Kollatschny, W. \& Bischoff, K. 2002, A{\&}A, 386, L19.

\bibitem[\protect\citeauthoryear{Kollatschny} {2003}]{Koll03} Kollatschny, W. 2003,  A{\&}A, 407, 461

\bibitem[\protect\citeauthoryear{La Mura et al.} {2007}]{Gio07} La Mura, G., Popovi\'c, L. \v C., Ciroi, S., Rafanelli, P., Ili\'c, D. 2007, ApJ, 671, 104L


\bibitem[\protect\citeauthoryear{La Mura et al.} {2009}]{lam09}	La Mura, G.; Mille, F. Di; Ciroi, S.; Popovi\'c, L. \v C.; Rafanelli, P., 2009. APJ, 639, 1437.

\bibitem[\protect\citeauthoryear{Livio \& Xu} {1997}]{Liv97} Livio, M. \& Xu, C. 1997, ApJ, 478, L63.

\bibitem[\protect\citeauthoryear{Marziani et al.} {2009}]{mar09}
	Marziani, P., Sulentic, J. W., Stirpe, G. M., Zamfir, S., Calvani, M. 2009,
A{\&}A, 495, 83

\bibitem[\protect\citeauthoryear{Murray \& Chiang}{1995}] {Mur95} Murray, N. \& Chiang, J. 1995, ApJ, 454, 105.

\bibitem[\protect\citeauthoryear{Murray \& Chiang}{1997}] {Mur97} Murray, N. \& Chiang, J. 1997, ApJ, 474, 91.

\bibitem[\protect\citeauthoryear{Murray \& Chiang}{1998}] {Mur98} Murray, N. \& Chiang, J. 1998, ApJ, 494, p.125.

\bibitem[\protect\citeauthoryear{Perez et al.} {1988}]{Perez88} Perez, E., Mediavilla, E., Penston, M. V., Tadhunter, C., Moles, M. 1988, MNRAS, 230, 353.

\bibitem[\protect\citeauthoryear{Popovi\'c et al.} {2009}]{Pop09} Popovi\'c, L. \v C.,  A. A. Smirnova, J. Kova\v cevi\'c, A. V. Moiseev, and V. L. Afanasiev, 2009, AJ, 137, 3548.

\bibitem[\protect\citeauthoryear{Popovi\'c et al.} {2008}]{Pop08} Popovi\'c, L. \v C., Bon, E., Gavrilovi\'c N., 2008, RevMexAA SC, 32, 99.

\bibitem[\protect\citeauthoryear{Popovi\'c et al.} {2004}]{Pop04} Popovi\'c, L. \v C., Mediavlilla, E.G., Bon, E., Ili\'c, D., 2004, A{\&}A, 423, 909.

\bibitem[\protect\citeauthoryear{Popovi\'c et al.} {2003}]{Pop03a} Popovi\'c, L. \v C., Mediavlilla, E.G., Bon, E., Stani\'c, N., Kubi\v cela, A., 2003, ApJ, 599, 185.

\bibitem[\protect\citeauthoryear{Popovi\'c et al.}  {2002}]{Pop02} Popovi\'c, L. \v C., Mediavlilla, E.G., Kubi\v cela, A., Jovanovi\'c, P., 2002, A\&{A}, 390, 473.

\bibitem[\protect\citeauthoryear{Popovi\'c et al.} {2001}]{Pop01} Popovi\'c, L. \v C., Stani\'c, N., Kubi\v cela, A., Bon, E. 2001,A\&{A}, 367, 780.

\bibitem[\protect\citeauthoryear{Proga \& Kallman} {2004}]{Proga04} Proga, D. and Kallman, T. R., 2004, ApJ, 616, 688


\bibitem[\protect\citeauthoryear{Rodr\'iguez-Ardila et al.} {1996}]{Rod96} Rodr\'iguez-Ardila, A., Pastoriza, M.G., Bica, E. 1996, ApJ, 463, 522.

\bibitem[\protect\citeauthoryear{Rokaki \& Boisson} {1999}]{RB99} Rokaki, E. \& Boisson, C. 1999, MNRAS 307, 41.

\bibitem[\protect\citeauthoryear{Romano et al.} {1996}]{Rom96} Romano, P, Zwitter, T., Calvani, M., Sulentic, J. 1996, MNRAS, 279, 165.

\bibitem[\protect\citeauthoryear{Smith et al.} {2005}]{Smith05} Smith, J. E., Robinson, A., Young, S., Axon, D. J., Corbett, Elizabeth A. 2005, MNRAS, 359, 846S.

\bibitem[\protect\citeauthoryear{Shapovalova et al.} {2004}]{Shap04} Shapovalova, A. I., Doroshenko, V. T., Bochkarev, N. G., Burenkov, A. N., Carrasco, L., Chavushyan, V. H., Collin, S., Valdes, J. R., Borisov, N., Dumont, A.-M., Vlasuyk, V. V., Chilingarian, I., Fioktistova, I. S., Martinez, O. M., 2004, A\&A, 422, 925

\bibitem[\protect\citeauthoryear{Shields et al.} {2000}]{Shields00} Shields, J.C., Rix, H.-W., McIntosh, D.H. et al. 2000, ApJ 534, L27.

\bibitem[\protect\citeauthoryear{Storchi-Bergmann et al.} {1997}]{Storchi97} Storchi-Bergmann, T.,  Eracleous, M., Ruiz, M.T., Livio, M., Wilson, A.S. \& Filippenko, A.V. 1997, ApJ, 489, 87.

\bibitem[\protect\citeauthoryear{Storchi-Bergmann et al.} {2003a}]{Berg03} Storchi-Bergmann, T., Nemmen, R., Eracleous, M.,  Halpern, J. P., Filippenko, A. V., Ruiz, M. T.,  Smith, R. C., Nagar, N. 2003a, ApJ, 598, 956.

\bibitem[\protect\citeauthoryear{Storchi-Bergmann et al.} {2003b}]{Stor03} Storchi-Bergmann, T.  de Silva, R.N., Eracleous, M. 2003b,  ASPC, 290, 155


\bibitem[\protect\citeauthoryear{Strateva et al.} {2003}]{Strateva03} Strateva, I.V., Strauss, M.A., Hao, L. et al. 2003, AJ, 126, 1720

\bibitem[\protect\citeauthoryear{Sulentic et al.} {2009}]{sul09} 
Sulentic, J. W., Marziani, P. \&  Zamfir, N.S. 2009, will appear in New. Astr. Rev.
doi:10.1016/j.newar.2009.06.001 

\bibitem[\protect\citeauthoryear{Sulentic et al.} {2000}]{Sul00} Sulentic, J. W., Marziani, P. \& Dultzin-Hacyan, D. 2000, ARAA 38, 521.


\bibitem[\protect\citeauthoryear{Urry \& Padovani} {1995}]{urpad95} Urry M. C. \& Padovani P. 1995, PASP 107, 803.


\bibitem[\protect\citeauthoryear{Wang et al.} {2003}]{Wang03} Wang, J.-M. Ho, L. C., Staubert, R. 2003, A\&{A}, 409, 887.
	
\bibitem[\protect\citeauthoryear{Wang et al.} {2005}]{wang05} Wang, T.-G., Dong, X.-B., Zhang, X.-G., Zhou, H.-Y., Wang, J.-X., Lu, Y.-J. 2005, ApJ, 625, L35 

\bibitem[\protect\citeauthoryear{Wills et al.} {2003}]{Wil93} Wills, B. J., Brotherton, M. S., Fang, D., Steidel, C. C., \& Sargent, W. L. W. 1993, ApJ, 415, 563
\end{thebibliography}
\end{document}